\documentclass[aps,
    twocolumn,
    amsmath,amssymb,
    raggedbottom,
    reprint,
    amsmath,
    amssymb,
    aps,
]{revtex4-2}

\usepackage[skip=5pt,font=small,labelfont=bf,
   justification=Justified,
   format=plain]{caption}
\usepackage{graphicx, physics}
\usepackage{dcolumn}
\usepackage{bm}
\usepackage{comment}
\usepackage{amsthm}
\usepackage{mathtools}
\usepackage[dvipsnames]{xcolor}
\usepackage{dcolumn}
\usepackage{natbib}
\usepackage{yfonts}
\usepackage{tikz}
\usepackage{braket}
\usepackage[export]{adjustbox}
\usepackage{verbatim}
\usepackage{parskip}
\usepackage{hyperref}
\usepackage{fontawesome5}
\usepackage{cleveref}
\usepackage{float}
\hypersetup{
	colorlinks=true,
	urlcolor=mymaroon,
	linkcolor=mymaroon,
	citecolor=mymaroon,
	pdftitle={},
	pdfauthor={},
	pdfdisplaydoctitle=true,
	pdfstartview=FitH
}
\usepackage{orcidlink}
\usepackage{bbm}
\usepackage{enumitem}

\crefname{figure}{fig.}{figs.}
\Crefname{figure}{Fig.}{Figs.}
\Crefname{equation}{Eq.}{Eqs.}

\def\equationautorefname~#1\null{(#1)\null}

\usetikzlibrary{calc}
\usetikzlibrary{arrows.meta, decorations.pathmorphing, patterns, decorations.pathreplacing, decorations.markings}
\usetikzlibrary{matrix}

\definecolor{color2}{rgb}{0.368417, 0.506779, 0.709798}
\definecolor{color3}{rgb}{0.880722, 0.611041, 0.142051}
\definecolor{color5}{rgb}{0.560181, 0.691569, 0.194885}
\definecolor{color1}{rgb}{0.922526, 0.385626, 0.209179}
\definecolor{color6}{rgb}{0.528488, 0.470624, 0.701351}
\definecolor{color4}{rgb}{0.772079, 0.431554, 0.102387}
\definecolor{mred}{rgb}{0.70, 0.20, 0.20}

\def \be {\begin{equation}}
\def \ee {\end{equation}}

\def \d {\mathrm{d}}
\def \leq {\leqslant}

\definecolor{mymaroon}{RGB}{128,0,0}
\definecolor{orcidlogocol}{named}{mymaroon}

\newcommand{\SOFIAtypeset}{%
  \adjustbox{valign=t,scale=0.7,raise=0.05em}{%
    \includegraphics[height=1em]{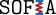}%
  }%
}
\DeclareRobustCommand{\SOFIA}{\texorpdfstring{\SOFIAtypeset}{SOFIA}}

\newcommand{\caveat}[1]{\hyperref[caveat:type#1]{Type~#1}}
\newcommand{\zenodorepo}{\href{https://doi.org/10.5281/zenodo.21298972}{\includegraphics[height=1.6ex]{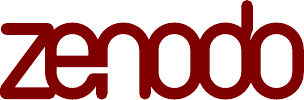}}}

\usepackage{iftex}
\ifpdf\usepackage{pdfrender}\fi
\definecolor{disckosky-blue}{rgb}{0.1216,0.3725,0.7490}
\definecolor{disckosky-red}{rgb}{0.7843,0.0627,0.1804}
\makeatletter
\newdimen\DK@cap
\newdimen\DK@x
\newsavebox\DK@box
\newcommand\DK@hidden[2]{%
  \ifpdf\rlap{\resizebox{#1}{!}{\textpdfrender{TextRenderingMode=Invisible}{#2}}}\fi}
\DeclareRobustCommand\DiscKosky{%
  \leavevmode
  \begingroup
    \edef\DK@test{\noexpand\in@{b}{\f@series}}\DK@test
    \ifin@ \expandafter\DK@full \else \expandafter\DK@light \fi
  \endgroup}
\newcommand\DK@full{%
  \DK@cap=\fontcharht\font`D\relax
  \ifdim\DK@cap=\z@ \DK@cap=.7em\fi
  \DK@x=\fontcharht\font`x\relax
  \ifdim\DK@x=\z@ \DK@x=1ex\fi
  \DK@cap=\dimexpr(\DK@cap+\DK@x*783/591)/2\relax
  \sbox\DK@box{\raisebox{-\dimexpr\DK@cap*192/783\relax}{%
    \includegraphics[height=\dimexpr\DK@cap*975/783\relax]%
      {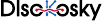}}}%
  \DK@hidden{\wd\DK@box}{DiscKosky}%
  \usebox\DK@box}
\newcommand\DK@light{%
  \DK@x=\fontcharht\font`x\relax
  \fontfamily{SourceCodePro-TLF}\fontseries{m}\fontshape{n}\selectfont
  \fontsize{\strip@pt\dimexpr\f@size pt*\DK@x/\fontcharht\font`x\relax}%
    {\f@baselineskip}\selectfont
  \pgfmathsetmacro\DK@em{\the\dimexpr1em\relax}%
  \pgfmathsetmacro\DK@cell{\the\fontcharwd\font`o/\DK@em}%
  \pgfmathsetmacro\DK@xc{-0.45*\DK@cell}%
  \def\DK@track{-0.05}%
  \pgfmathsetmacro\DK@box{\DK@cell+2*\DK@track}%
  \pgfmathsetmacro\DK@xo{\DK@box+0.5*\DK@cell}%
  \pgfmathsetmacro\DK@yw{0.5*\the\fontcharht\font`x/\DK@em}%
  \pgfmathsetmacro\DK@yt{\the\fontcharht\font`K/\DK@em-0.048}%
  \pgfmathsetmacro\DK@P{(\DK@xo-\DK@xc)/2.5}%
  \pgfmathsetmacro\DK@xs{\DK@xc+\DK@P}%
  \pgfmathsetmacro\DK@xr{\DK@xs+0.5*\DK@cell}%
  \def\DK@wave##1{\DK@yw+0.036*sin(360*(##1-\DK@xc)/\DK@P)}%
  \begin{tikzpicture}[baseline=0pt, x=1em, y=1em, inner sep=0pt, outer sep=0pt]
    \node[anchor=base west] (Disc) at (0,0)
      {D\kern\DK@track em i\kern\DK@track em s\kern\DK@track em c};
    \begin{scope}[shift={(Disc.base east)}]
      \node[anchor=base west] at (0,0) {\DK@hidden{\DK@box em}{K}};
      \begin{scope}[even odd rule]
        \clip (-1,-1) rectangle (4,1)
          plot[domain=\DK@xc:\DK@xo, samples=41, variable=\t]
            (\t, {\DK@wave{\t}+0.043})
          -- plot[domain=\DK@xo:\DK@xc, samples=41, variable=\t]
            (\t, {\DK@wave{\t}-0.043}) -- cycle;
        \node[anchor=base west] (osky) at (\DK@box,0)
          {o\kern\DK@track em s\kern\DK@track em k\kern\DK@track em y};
      \end{scope}
      \begin{scope}[disckosky-blue, line width=0.05em, line cap=round,
                    line join=round]
        \draw (\DK@xs,0.048) -- (\DK@xs,\DK@yt)
          (\DK@xr,\DK@yt) -- (\DK@xs,\DK@yw) -- (\DK@xr,0.048);
        \fill (\DK@xs,0.048) circle[radius=0.048]
          (\DK@xs,\DK@yt) circle[radius=0.048]
          (\DK@xr,0.048) circle[radius=0.048]
          (\DK@xr,\DK@yt) circle[radius=0.048];
      \end{scope}
      \begin{scope}[disckosky-red]
        \draw[line width=0.05em, line cap=round]
          plot[domain=\DK@xc:\DK@xo, samples=61, smooth, variable=\t]
            (\t, {\DK@wave{\t}});
        \fill (\DK@xc,\DK@yw) circle[radius=0.058]
          (\DK@xo,\DK@yw) circle[radius=0.058];
      \end{scope}
    \end{scope}
    \pgfresetboundingbox
    \useasboundingbox (Disc.base west |- osky.south) rectangle (osky.north east);
  \end{tikzpicture}}
\usepackage{tcolorbox}
\usepackage{colortbl} 
\usepackage{listings}
\usepackage{setspace} 
\definecolor{identifiercolor}{rgb}{.4,.6,.56}
\definecolor{stringcolor}{gray}{0.5}
\definecolor{inactivecolor}{rgb}{0.2,0.2,0.2}

\definecolor{light-gray}{gray}{0.95}

\definecolor{gh-fg}{HTML}{000000}
\definecolor{gh-comment}{HTML}{007ACC}
\definecolor{gh-keyword}{HTML}{E60000}
\definecolor{gh-func}{HTML}{FF7F00}
\definecolor{gh-string}{HTML}{999999} 
\definecolor{gh-number}{HTML}{000000}
\definecolor{gh-linenum}{HTML}{666666}
\definecolor{gh-frame}{HTML}{BBBBBB}
\definecolor{disckosky-purple}{HTML}{22774d}

\begin{document}

\newsavebox\ttlbox
\savebox\ttlbox{\parbox{\textwidth}{\centering
  {
  \large\textbf{Landau's Leviathans}
  }\\[2pt]
  {\emph{Singularities from Euler-Characteristic Drops over Finite Fields}}}}
\title{\texorpdfstring{\usebox{\ttlbox}}{Landau's Leviathans: Singularities from Euler-Characteristic Drops over Finite Fields}}
\author{Vsevolod Chestnov$^{a,\orcidlink{0000-0001-7067-0315}}$}
\author{Giulio Crisanti$^{b,\orcidlink{0009-0009-3053-2394}}$}
\author{Mathieu Giroux$^{c,d,\orcidlink{0000-0002-2672-634X}}$}%
\affiliation{$^a$Mathematical Institute, University of Oxford, OX2 6GG, UK}
\affiliation{$^b$Higgs Centre for Theoretical Physics, University of Edinburgh, James Clerk Maxwell Building, Peter Guthrie Tait Road, Edinburgh, EH9 3FD, UK}
\affiliation{%
$^c$Center for Theoretical Physics, Department of Physics, 
Columbia University, Pupin Hall,
538 West 120th Street, New York, NY 10027, USA}
\affiliation{$^d$Institute for Advanced Study, Einstein Drive, Princeton, NJ 08540, USA}

\begin{abstract}
\emph{Abstract.} We present a new method together with a proof-of-concept implementation for determining the Landau singularities of Feynman integrals, read off directly from where the Euler characteristic of the associated integral drops. Working over finite fields makes the requisite elimination tractable for multi-scale integrals at the multi-loop frontier. The algorithm returns the genuine and complete set of singularities, subject to a set of conditions which are practically testable. We apply these methods to classes of Feynman integrals beyond the reach of current tools, including non-planar six-point diagrams at two loops, as well as a fully massive three-loop envelope graph. Several of the newly found singularities, both in $d$- and 4-dimensional external kinematics, are of unexpected complexity when compared to previously known singularities for these examples.
\end{abstract}

\makeatletter\let\mtl@smash\smash\renewcommand{\smash}[1]{#1}\makeatother
\maketitle
\makeatletter\let\smash\mtl@smash\makeatother

\section{Introduction}

Modern high-precision phenomenology at colliders and gravitational-wave observatories relies on our ability to compute scattering amplitudes to ever higher loop orders. A lesson of the past decade is that the difficulty of these computations is controlled, to a large extent, by our understanding of the analytic structure of the underlying Feynman integrals: the \emph{locations} of branch points and discontinuities provide powerful constraints on their functional forms, often at much lower cost than head-on integration.

The locations of these branch points and discontinuities are not arbitrary. They arise where the integration contour is pinched between colliding singularities of the integrand, a mechanism that was recognised long ago \cite{Landau:1959fi,10.1143/PTP.22.128,Cutkosky:1960sp,Collins:2020euz}. Together, they form the \emph{set of Landau singularities} (or simply the \emph{Landau locus}) and delineate the boundary of analyticity of the integral.

However, for these data to be maximally useful, one would like to obtain them \emph{before} committing to a full computation of the integrals, which grows prohibitively expensive with the number of loops and legs. A productive idea has been to invert the logic and \emph{leverage} the Landau locus: once known, it provides input to the symbol alphabet, to integration-by-parts and differential-equation methods \cite{Hannesdottir:2021kpd,Dennen:2015bet,Dlapa:2023cvx,Coro:2025kha,Correia:2025yao} as well as direct evaluation strategies \cite{Panzer:2014caa,Giroux:2026tgd}, and it seeds bootstrap and ansatz-based reconstructions of the individual Feynman integrals \cite{Hannesdottir:2024hke,Barrera:2025uin} or even the amplitude itself \cite{Bern:1994cg,Caron-Huot:2020bkp,Correia:2020xtr,Correia:2021etg,Carrolo:2025agz}.

Determining the Landau locus more systematically beyond one loop has only recently become feasible through several distinct frameworks. Building on the modern reformulation of \cite{Mizera:2021icv,Klausen:2021yrt}, the principal Landau determinant (PLD) recast the problem in computational algebraic geometry and made the extraction of the Landau locus practical for realistic multi-scale processes \cite{Fevola:2023kaw,Fevola:2023fzn}. A separate line extracts singularities from stratifications of the relevant varieties \cite{Helmer:2024wax,Helmer:2025yuf,Helmer:2025ljj,Telen:2024sep}.
The automated \textsc{Mathematica} program \SOFIA{}, working within yet another framework \cite{Panzer:2014caa,Caron-Huot:2024brh}, organises the analysis by sectors and, among the above, currently offers the broadest practical coverage of examples relevant, e.g., for precision Standard Model physics.

However, two limitations persist. First, all of these methods become computationally prohibitive on certain classes of graphs. For example, some multi-loop (near) maximally connected diagrams (e.g., \cref{fig:new_applications}) represent a sharp class of \emph{bottleneck diagrams} that mark the current computational frontier, where new and complementary methods are needed.

Second, many of these methods return a \emph{superset} of the genuine Landau locus, whose spurious factors must be removed by hand, or, worse, a \emph{subset} that misses genuine singularities altogether and is therefore incomplete for practical amplitude computations.

In this Letter we propose a route that addresses both issues simultaneously. The key object is a number known as the (absolute value of the) \emph{Euler characteristic} $\chi(\boldsymbol{s})$ associated to the Feynman integral, or more generally to any parametric (Euler) integral in Lee--Pomeransky form \cite{Lee:2013hzt}. Physically, it counts the number of master integrals \cite{Lee:2013hzt,Frellesvig:2019kgj}, which in turn equals the number of critical points of the associated parametric integral \cite{Matsubara-Heo:2025lrq}.

Euler characteristics have been used extensively to \emph{test} whether a candidate singularity is genuine \cite{Fevola:2023fzn}. In this work we turn this idea around and propose to use $\chi$ to \emph{find} the locus directly (see also \cite{Telen:2024sep}). In this picture, the Landau singularities are precisely the locations in kinematic space where $\chi$ drops, i.e., where some critical points ``escape'' to a boundary of the parametric space.

Detecting where this happens is made efficient by working with finite field arithmetic, which renders the required algebra tractable even for the many-parameter systems generated by cutting-edge multi-loop and multi-scale Feynman integrals. In practice, we use the package SP$\mathbb{Q}$R \cite{Chestnov:2025svg} for polynomial algebra, which leverages \texttt{FiniteFlow} \cite{Peraro:2016wsq,Peraro:2019svx} as its primary back end.

Because this method pinpoints the locus from the critical points that are genuinely lost, this approach returns the strict set of genuine singularities directly rather than a superset. The computational framework behind this construction is simple, and we accompany this Letter with a proof-of-concept \textsc{Mathematica} implementation showcasing its most important features.

For practical computational purposes, one often works ``sector-by-sector'', which trades some mathematical rigour for computational speed. Nevertheless, the full set of singularities can still be found in the vast majority of cases, subject to a set of conditions which are practically testable.

After describing the construction in detail, we put the method to the test on three bottleneck diagrams beyond the reach of existing tools: the two hardest non-planar six-point diagrams at two loops, as well as the fully massive four-scale three-loop non-planar envelope, all shown in \cref{fig:new_applications}. The resulting singularity lists and our proof-of-concept \textsc{Mathematica} implementation are collected in a \zenodorepo{} repository accessible at 
\begin{center}
\url{https://doi.org/10.5281/zenodo.21298972}.
\end{center}

\section{Method}\label{sec:method}

\textbf{Euler Characteristics.}\quad Our strategy for locating the singularities of Feynman integrals revolves around computing the Euler characteristic $\chi$ as a function of kinematic variables $\boldsymbol{s}$.
To this end, we consider a Feynman integral with $E$ internal edges in Lee--Pomeransky \cite{Lee:2013hzt} representation in $d$ spacetime dimensions
\begin{equation}\label{eq:reg_I}
    I_{\boldsymbol{\nu}}(\boldsymbol{s}) \propto \int_0^\infty \boldsymbol{x}^{\boldsymbol{\nu}} \mathcal{G}(\boldsymbol{x},\boldsymbol{s})^{-d/2} \frac{\dd\boldsymbol{x}}{\boldsymbol{x}}\qquad (\boldsymbol{\nu},d)\in\mathbb{C}^{E+1}\,.
\end{equation}
Here, $\boldsymbol{x}^{\boldsymbol{\nu}}\equiv x_1^{\nu_1}\ldots x_E^{\nu_E}$ and an irrelevant kinematically independent prefactor has been omitted from our discussion. Although we focus on the Lee--Pomeransky representation in this work, the method can be applied to other parametric representations, such as standard Baikov \cite{Baikov:1996cd,Baikov:2005nv,Frellesvig:2024ymq}, as well as the broader class of Euler integrals (see e.g. the review \cite{Matsubara-Heo:2023ylc}). The Euler characteristic associated to $I_{\boldsymbol{\nu}}(\boldsymbol{s})$ is then $\chi(X_{\boldsymbol{s}})$, where 
\begin{equation}
    X_{\boldsymbol{s}} =\mathbb{C}^E\setminus V(\boldsymbol{x} \cdot\mathcal{G}(\boldsymbol{x},\boldsymbol{s}))\,,
\end{equation}
and $V(F)$ denotes the vanishing set given by $F=0$. Up to an overall irrelevant sign, $\chi(X_{\boldsymbol{s}})$ can be practically computed \cite{Huh:2013} as
\begin{equation}\label{eq:chi1}
\begin{aligned}
    \omega &= \mathrm{d}\log\bigl(\boldsymbol{x}^{\boldsymbol{\nu}}\mathcal{G}^{-d/2}(\boldsymbol{x},\boldsymbol{s})\bigr)\,,\\
    \chi(X_{\boldsymbol{s}}) &=(-1)^E \times\left(\# \text{ of solutions to} \,\, \omega = 0\right)\,.
\end{aligned}
\end{equation}
Physically, $(-1)^E \chi(X_{\boldsymbol{s}})$ counts the number of master integrals of the \emph{regulated} Feynman integral family $I_{\boldsymbol{\nu}}(\boldsymbol{s})$ \cite{Bitoun:2017nre,Frellesvig:2019uqt}.

\Cref{eq:chi1} is equivalent to the system of polynomial equations (referred to as an \emph{ideal}), given by
\begin{equation}\label{eq:chi_ideal_reg}
\hspace{-0.293cm}
    \mathcal{I}\equiv\left\langle
    \nu_1\mathcal{G}-\frac{d}{2}x_1 \partial_1\mathcal{G}
    ,{\ldots},
    \nu_E \mathcal{G}-\frac{d}{2} x_E \partial_E\mathcal{G},
    1{-}x_0\boldsymbol{x}
    \mathcal{G}
    \right\rangle,
\end{equation}
where the final generator, together with the auxiliary variable $x_0$, removes any solutions lying on the hypersurface $x_1\cdots x_E\,\mathcal{G}=0$. Equivalently, $\mathcal{I}$ describes the critical points on the variety $X_{\boldsymbol{s}}$, whose boundary components are $\{x_i=0\}\cup\{x_i=\infty\}\cup\{\mathcal{G}=0\}$. For generic $(\boldsymbol{\nu},d)$ the ideal $\mathcal{I}$ is \textit{always} zero-dimensional \cite{Franecki:1999gcm,Huh:2013}, so its vanishing (or solution) set, denoted as $V(\mathcal{I})$, consists of finitely many isolated \emph{critical points}. The number of such points (with multiplicity) is known as the \emph{degree} of the ideal
\begin{equation}\label{eq:ChiDeg}
    (-1)^E\chi(X_{\boldsymbol{s}}) = \deg(\mathcal{I}(\boldsymbol{s}))\,.
\end{equation}

For generic kinematics $\boldsymbol{s}$, the value of $\chi(X_{\boldsymbol{s}})$ is constant. On a special locus $\boldsymbol{s}^*$, however, the signed Euler characteristic may drop discontinuously to a lower value \cite{Fevola:2023fzn,Matsubara-Heo:2025lrq,Helmer:2025yuf,Helmer:2025ljj}:
\begin{equation}\label{eq:euler_chi_drop}
    (-1)^E \chi(X_{\boldsymbol{s}^*}) < (-1)^E \chi(X_{\boldsymbol{s}}) \,.
\end{equation}
Since $\chi$ is a topological invariant, such a drop signals a change in $X_{\boldsymbol{s}}$'s topology, occurring precisely where a new singularity develops at a point $\boldsymbol{s}^*$ in kinematic space. These special locations are therefore in one-to-one correspondence with the branch points of $I_{\boldsymbol{\nu}}(\boldsymbol{s})$: writing $l(\boldsymbol{s})=l_1(\boldsymbol{s})\cdots l_N(\boldsymbol{s})$ for the product of all singularities of the Feynman integral, \cref{eq:euler_chi_drop} holds precisely when $l(\boldsymbol{s}^*)=0$. From now on, we will refer to these singularities of \cref{eq:reg_I} as \emph{Landau singularities} \cite{Fevola:2023fzn,Matsubara-Heo:2025lrq}.

\Cref{eq:euler_chi_drop} has been used extensively to \textit{test} whether a given candidate Landau singularity is genuine or not \cite{Fevola:2023kaw,Fevola:2023fzn}. In this work, we present a new method that instead uses $\chi(\boldsymbol{s})$ to \textit{find} the $l(\boldsymbol{s})$ associated to \cref{eq:reg_I}.

To this end, we let $p$ be an element of the solution set to \cref{eq:chi_ideal_reg}, $p \in V(\mathcal{I})$, and let $|p(\boldsymbol{s})|$ be its distance from the origin in $X_{\boldsymbol{s}}$. Crucially, as the number of solutions to \cref{eq:chi_ideal_reg} degenerates at $\boldsymbol{s} = \boldsymbol{s}^*$, at least one solution point must run off to infinity as $\boldsymbol{s}\to\boldsymbol{s}^*$,
\begin{equation}
    \lim_{\boldsymbol{s} \to \boldsymbol{s}^*} |p(\boldsymbol{s})| = \infty \quad \text{for some } p \in V(\mathcal{I}) \,.
\end{equation}
This can be seen as follows. For generic $(\boldsymbol{\nu},d)$, $V(\mathcal{I})$ remains a finite set of isolated points for any value of $\boldsymbol{s}$. Their number (with multiplicity) can therefore drop only when a solution reaches a boundary of $X_{\boldsymbol{s}}$.

\begin{figure}
    \centering
    \includegraphics[scale=0.92]{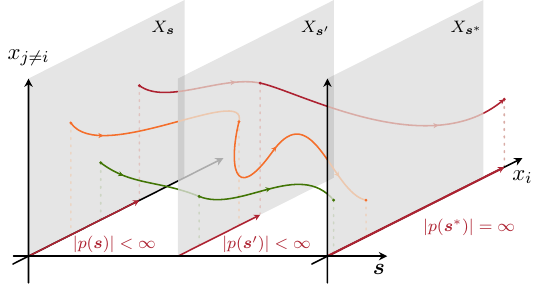}
    \caption{
        A schematic of the projection strategy. As the kinematics change towards a singular configuration, a critical point (red) moves to infinity. This behaviour manifests itself by (at least) one projection onto a coordinate axis $x_i$ for which $|p|$ takes finite values at $\boldsymbol{s}$ and $\boldsymbol{s}'$, but diverges at $\boldsymbol{s}^*$. This is captured by $\chi(X_{\boldsymbol{s}})=\chi(X_{\boldsymbol{s}'})=3$ and $\chi(X_{\boldsymbol{s}^*})=2$.
    }
    \label{fig:crit_point_projection}
\end{figure}

In the coordinates $(x_0,\ldots,x_E)$, every such escape sends some coordinate to infinity: in particular, a point reaching $\{x_i=0\}$ or $\{\mathcal{G}=0\}$ makes $x_0=1/(x_1\cdots x_E\,\mathcal{G})\to\infty$, while $\{x_{i>0}=\infty\}$ is direct. The proposed strategy is thus to track the values of $\boldsymbol{s}$ for which this singular behaviour is observed. This is illustrated in \cref{fig:crit_point_projection}.

\textbf{Ideal Projections.}\quad Identifying when a solution diverges to infinity for a multivariate system of polynomial equations is at first sight a difficult problem. It is thus illustrative to look at the univariate case first, where ideals are generated by a single polynomial,
\begin{equation}\label{eq:ideal_univ_ex}
    \mathcal{I}_{\text{univ}}(\boldsymbol{a}) = \left\langle a_0 + a_1 x + \cdots+a_{m-1}x^{m-1} + a_m x^m \right\rangle \,,
\end{equation}
with bounded coefficients $a_0,\ldots,a_m$. Such a polynomial has $m$ roots (counted with multiplicity). The only way for this number to drop is if the degree of the polynomial decreases, which can only occur if $a_m = 0$. Indeed, in the limit $a_m \to 0$, at least one root of the polynomial above escapes to infinity. In the context of Feynman integrals, $a_m$ would thus correspond to a Landau singularity (or, more generally, a product of them).

In a multivariate setting such as \cref{eq:chi_ideal_reg}, there is no direct analogue of the univariate formula in \cref{eq:ideal_univ_ex} from which to read off the relevant coefficients. Nevertheless, one can reduce the problem to the univariate case by projecting the solution set onto one coordinate $x_i$ at a time, before repeating the univariate analysis described above. This can be achieved by computing \textit{elimination ideals}:
\begin{equation}
\mathcal{I}_{\text{elim}}^{(i)}
    \equiv\mathcal{I}\cap\mathbb{Q}[\boldsymbol{s}, x_i]\,.
\end{equation}
Intuitively $\mathcal{I}_{\text{elim}}^{(i)}$ consists of all polynomial relations in $\mathcal{I}$ involving \textit{only} the variable $x_i$. Since $\mathcal{I}$ is zero-dimensional, $\mathcal{I}_{\text{elim}}^{(i)}$ is generated by a single univariate polynomial whose roots are precisely the $x_i$-coordinates of the solution points $p$,
\begin{equation}
    \mathcal{I}_{\text{elim}}^{(i)}
    =
    \left\langle
    c^{(i)}_0(\boldsymbol{s}) + \cdots + c^{(i)}_m(\boldsymbol{s}) x_i^m
    \right\rangle\,.
    \label{eq:elim}
\end{equation}
Exactly as in the univariate example, an $x_i$-coordinate can diverge only when the leading coefficient vanishes, so $c^{(i)}_m(\boldsymbol{s})$ contains a subset of the Landau singularities of the Feynman integral. By iterating the procedure over all $i \in \{0,\ldots,E\}$, \textit{all} singularities will be captured.

Computing the elimination ideal in \cref{eq:elim} requires computing \emph{Gr\"obner bases} and polynomial reductions modulo them, which can become prohibitively expensive. We sidestep much of this cost by working over finite fields: the recently released linear-algebraic reduction package SP$\mathbb{Q}$R~\cite{Chestnov:2025svg}, through \texttt{FiniteFlow}~\cite{Peraro:2019svx}, makes such projections feasible even for the many (kinematic) scale polynomial systems generated by Feynman integrals.

\textbf{Decomposition in Sectors.}\quad In practice, the computation of $\chi$ can be greatly simplified by stratifying the problem into \emph{sectors}. A sector is labelled by a tuple $\mathcal{S}\in\{0,1\}^E$ whose $i$-th entry is $1$ if propagator $i$ is kept and $0$ if it is contracted, so that $|\mathcal{S}|$ counts the total number of propagators and $\mathcal{S}=(1,\ldots,1)$ is the top sector. For each sector, let
\begin{equation}\label{eq:sector_X_space}
    X_{\boldsymbol{s}}^{\mathcal{S}} \equiv \mathbb{C}^{|\mathcal{S}|} \setminus V(\mathcal{G}_{\mathcal{S}})\,,
\end{equation}
where the polynomial $\mathcal{G}_\mathcal{S}\equiv\mathcal{G}\big|_{x_i=0\;\forall i\,:\,\mathcal{S}_i=0}$ is the Lee-Pomeransky polynomial of the given (sub)sector. By using the inclusion--exclusion property of $\chi$, it can be shown that \cite{Bitoun:2017nre}
\begin{equation}\label{eq:inclusion_exclusion}
    (-1)^E\chi(X_{\boldsymbol{s}}) = \sum_{\mathcal{S}\in\text{sectors}}(-1)^{|\mathcal{S}|} \, \chi(X_{\boldsymbol{s}}^{\mathcal{S}})\,.
\end{equation}
Crucially, each $X_{\boldsymbol{s}}^{\mathcal{S}}$ now \textit{includes} the coordinate hyperplanes that are excluded from the original space $X_{\boldsymbol{s}}$. \Cref{eq:inclusion_exclusion} can thus be interpreted as the $\boldsymbol{\nu} \to 0$ specialisation of the \textit{regulated} Feynman integral to the \textit{unregulated} case \cite{Lee:2013hzt}. When this limit is well behaved, $(-1)^{|\mathcal{S}|}\chi(X_{\boldsymbol{s}}^{\mathcal{S}})$ can then be interpreted as counting the number of master integrals in the given sector $\mathcal{S}$.
By defining
\begin{equation}\label{eq:chi_ideal_unreg}
    \mathcal{J}_\mathcal{S} \equiv \langle \partial_1 \mathcal{G}_\mathcal{S},\,\ldots,\,\partial_{|\mathcal{S}|}\mathcal{G}_\mathcal{S},\,1-x_0\,\mathcal{G}_\mathcal{S} \rangle \,,
\end{equation}
where the index now labels the non-zero sector variables.
These Euler characteristics can then very often be computed in an identical manner to the $\boldsymbol{\nu}\neq 0$ setting \cite{Lee:2013hzt}:
\begin{equation}\label{eq:euler_chi_sector_sum2}
    \chi(X_{\boldsymbol{s}}^{\mathcal{S}}) \stackrel{*}{=} (-1)^{|\mathcal{S}|}\deg(\mathcal{J}_\mathcal{S})\,,
\end{equation}
and thus
\begin{equation}\label{eq:euler_chi_sector_sum}
    (-1)^E\chi(X_{\boldsymbol{s}})=\deg(\mathcal{I}) \stackrel{*}{=} \sum_{\mathcal{S}\in\text{sectors}}\deg(\mathcal{J}_\mathcal{S})\,.
\end{equation}
where $\stackrel{*}{=}$ denotes equality barring rare exceptions (see Appendix \ref{section:caveats}). Finding singularities then reduces to checking whether any individual $|\chi(X_{\boldsymbol{s}}^{\mathcal{S}})|$, and thus $\deg(\mathcal{J}_\mathcal{S})$, drops from its generic value, using the same elimination strategy applied to each $\mathcal{J}_\mathcal{S}$.

This decomposition provides substantial computational benefit, splitting one large computation into many small ones. However, it is optional: one may always work directly with the full ideal in \cref{eq:chi_ideal_reg}, at the price of a higher computational cost.

Importantly, the $\boldsymbol{\nu}=0$ specialisation may carry subtleties, which manifest themselves as \cref{eq:euler_chi_sector_sum2,eq:euler_chi_sector_sum} failing to hold, either generically or on special kinematic loci. We discuss these in Appendix \ref{section:caveats}, where we also provide efficient diagnostic tests that detect such situations.

\section{A Working Example}
\textbf{Two-Loop Equal-Mass Sunrise.}\quad To illustrate the methods discussed in \cref{sec:method}, it is useful to work through a simple example. To this end, we consider
\begin{align}\label{eq:sunriseG}
    &\hspace{2cm}\includegraphics[valign=c]{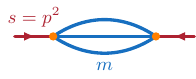}\notag\\
\mathcal{G}_{\text{sun}}&(\boldsymbol{x},s,m) =
x_1 x_2
+ x_1 x_3
+ x_2 x_3
+ s\,x_1 x_2 x_3
\\&
-m^2\,
(x_1+x_2+x_3)(x_2x_3+x_1x_2+x_1x_3)
\notag\,,
\end{align}
namely a two-loop equal-mass sunrise diagram. Using \cref{eq:chi_ideal_unreg}, we find that $\deg(\mathcal{J}_{\text{sun}}^{\mathcal{S}})=-\chi(X_{\boldsymbol{s}}^{\mathcal{S}})=4$,
where $\mathcal{S}=(1,1,1)$ is the top sector including all three propagators. This is in agreement with the well-known fact that there are four master integrals (without accounting for symmetries) in the top sector of this diagram. To find the Landau singularities, one proceeds by projecting down $\mathcal{J}_{\text{sun}}$ onto each coordinate axis. For $x_0$ this results in a quadratic polynomial:
\begin{equation}
    \mathcal{J}_{\text{sun}} \cap \mathbb{Q}[\boldsymbol{s}, x_0] = \left\langle c_0^{(0)}+c_1^{(0)}x_0+c_2^{(0)}x_0^2 \right\rangle\,.
\end{equation}
Thus, the signed Euler characteristic drops when $c_2^{(0)}=0$. Indeed, an explicit computation shows that
\begin{equation}
    c_2^{(0)}=16\,s\,,
\end{equation}
which produces the Landau singularity $\{s\}$. The procedure can now be repeated for the other coordinates. For $x_1$ the projection results in a cubic polynomial
\begin{equation}
    \label{eq:sunrise_ideal}
    \mathcal{J}_{\text{sun}} \cap \mathbb{Q}[\boldsymbol{s}, x_1] = \left\langle c_0^{(1)}+\cdots+c_3^{(1)}x_1^3 \right\rangle\,,
\end{equation}
where explicitly the top coefficient is given by
\begin{equation}
c_3^{(1)}=9\, m^2 \left(m^2-s\right)^2 \left(9 m^2-s\right)\,.
\end{equation}
Thus, $\{m^2,s-m^2,s-9m^2\}$ are identified as further Landau singularities. Repeating this procedure for the remaining two coordinates results in no new factors. By combining the singularities from all projections the full set is thus given by $\boldsymbol{l}(s,m)=\{s,m^2,s-m^2,s-9m^2\}$, reproducing the well-known standard result \cite{Mizera:2021icv,Laporta:2004rb}.

This result, and many others, can automatically be produced with the proof-of-concept \textsc{Mathematica} routine \texttt{SPQRLandau} available from the \zenodorepo{} repository:
\noindent\begin{minipage}{\linewidth}
\begin{lstlisting}
G = x1 x2 + x1 x3 + x2 x3 + s x1 x2 x3
    - m^2 (x1 + x2 + x3)(x2 x3 + x1 x2 + x1 x3);
vars = {x1, x2, x3};
SPQRLandau[G, vars] (* Out: {m^2,s,s-9m^2,s-m^2} *)
\end{lstlisting}
\end{minipage}

\begin{figure*}[t]
\noindent{\color{gray!50}\rule{\textwidth}{0.3pt}}\par
\vspace{0.15\baselineskip}
\centering
  \begin{minipage}[b]{0.27\textwidth}\centering
    \includegraphics[max width=\linewidth]{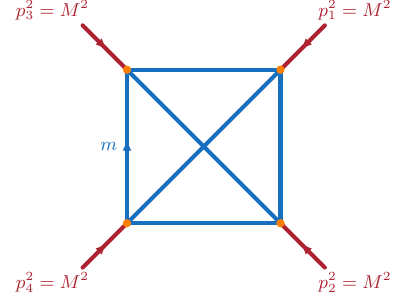}\\[2pt]%
    {\footnotesize (a)}\\[1pt]{\small $8$ new, len. $653$, deg. $21$}
  \end{minipage}\hfill
  \begin{minipage}[b]{0.27\textwidth}\centering
    \includegraphics[max width=\linewidth]{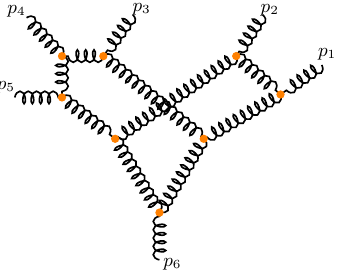}\\[2pt]%
    {\footnotesize (b)}\\[1pt]{\small $34$ new, len. $5\,543$, deg. $12$}
  \end{minipage}\hfill
  \begin{minipage}[b]{0.27\textwidth}\centering
    \includegraphics[max width=\linewidth]{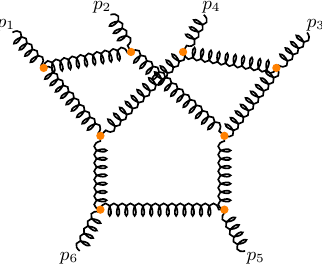}\\[2pt]%
    {\footnotesize (c)}\\[1pt]{\small $45$ new, len. $57\,571\,387$, deg. $36$}
  \end{minipage}
    \caption{(a) fully massive envelope, (b,c) non-planar six-point two-loop graphs. Below each graph: the number of new singularities found by our method (relative to \cite{PLDdata}), length and total degree (in $\text{mass}^2$) of the largest newly found singularity (Leviathan). All singularities except the largest one of (c) were generated on an M1 MacBook Pro with 32 GB of RAM\@. The largest singularity of (c) was computed on an AMD EPYC 9575F 64-Core cluster and required approximately 100 GB of memory.
  \label{fig:new_applications}
  }
\vspace{-0.3\baselineskip}
\noindent{\color{gray!50}\rule{\textwidth}{0.3pt}}
\end{figure*}

{\section{New Results}\label{sec:CE}}
The new method of \cref{sec:method} is especially suited to the cutting-edge \emph{bottleneck diagrams}, namely (near) maximally connected graphs. \Cref{fig:new_applications} shows three such examples. Their Landau singularities have been studied in the past, but the full set had remained beyond the reach of all currently available tools. In fact, every method we are aware of (e.g. \cite{Fevola:2023fzn,Correia:2025yao,Helmer:2024wax,Telen:2024sep}) either stalls or produces incomplete lists of singularities for these examples.

Using the method introduced above, all known singularities of \cite{PLDdata} were reproduced and a new set of singularities was resolved for all three families.

For each of the three families the diagnostic tests of Appendix \ref{section:caveats} were run. A small fraction of sectors (for example, 8\% for diagram (b)) triggered the specialised degeneracy of \cref{eq:degenerate_factoring}. These sectors are simple and their singularities could be computed directly by \SOFIA{} \cite{Correia:2025yao}, while the sectors accessible only through the projection pass all diagnostic checks. To the best of our understanding therefore, no genuine singularity has been missed by our analysis, and we regard these lists as \textit{complete}. The full set of singularities can be found in the \zenodorepo{} repository.

Perhaps one of the most surprising things about the additional sets of singularities obtained for these diagrams is the remarkable size and high polynomial degree of some of them, as documented in \cref{fig:new_applications} \footnote{%
We furthermore independently verified the drop in the number of master integrals on these singularities using IBP identities~\cite{Tkachov:1981wb,Chetyrkin:1981qh} on fully numerical slices, generated with the private \textsc{Mathematica} package~\textsc{FFIntRed} by Tiziano Peraro and solved by a variant of the Laporta algorithm~\cite{Laporta:2001dd} based on the finite-field arithmetic and functional reconstruction framework of~\textsc{FiniteFlow}~\cite{Peraro:2019svx}.
}.

We note that \texttt{SPQRLandau} is intended as a proof-of-concept function and is not sufficiently optimised to tackle the diagrams of \cref{fig:new_applications}. These integrals required a manual optimisation of the finite-field sampling pipeline. A full automation of this procedure is left to future work.

\textbf{Four-Dimensional Kinematics.}\quad The above results are computed assuming $d$-dimensional external kinematics, both to stress-test the method and to benchmark against PLD \cite{Fevola:2023fzn,Fevola:2023kaw} and \SOFIA{} \cite{Correia:2025yao}, which operate by default under this assumption. The singularity lists can be restricted to four-dimensional external kinematics \cite{tHooft:1972tcz}, using the parametrisation of \cref{section:4d}. We find that all restricted singularities are genuine (see the discussion around \cref{eq:4d_prefactor} for details). Both $d$- and four-dimensional results are listed in the \zenodorepo{} repository \footnote{%
    We also cross-checked the corresponding Euler characteristics for each sector using IBP identities, following the strategy of~\cite{Henn:2024ngj}.
}.

\section{Discussion}
In this Letter we have introduced a new method for determining the Landau singularities of a Feynman integral. The method extracts the entire Landau locus directly by tracking the kinematic configurations at which the Euler characteristic drops, that is, where critical points escape to infinity. Working over finite fields renders the underlying elimination tractable even for many-scale, multi-loop integrals \cite{Chestnov:2025svg}. The output, subject to verified diagnostic checks, is the strict set of singularities rather than a superset. For complicated polylogarithmic Feynman integrals at the precision frontier, working with the exact set could dramatically reduce the effort needed to construct symbol letters from candidate singularities \cite{repoEffortless}.

We have illustrated the method on bottleneck diagrams beyond the reach of existing tools, namely the two hardest non-planar six-point two-loop graphs and the non-planar massive envelope in \cref{fig:new_applications}. We provide a proof-of-concept implementation together with an extensive cross-check against \SOFIA{} \cite{Correia:2025yao}. Although suppressed in colour relative to their recently computed planar counterparts \cite{Abreu:2024fei,Henn:2025xrc,Liu:2026hdp}, the non-planar contributions to the six-gluon amplitude in \cref{fig:new_applications} (b,c) should soon be tackled by the community. The singularities found in this work, together with their four-dimensional restriction discussed in \cref{sec:CE}, should be essential to that effort.

Several directions invite further work. On the algorithmic side, interfacing the elimination step with dedicated solvers such as \texttt{msolve} \cite{msolve} should extend the reach to still more complex topologies. A finer understanding of the factored structure of the projected polynomials, including the multiplicities with which individual singularities appear, would further improve performance. It would also be worthwhile to evaluate the Euler characteristic in alternative representations, such as the loop-by-loop Baikov representation \cite{Frellesvig:2017aai,Frellesvig:2024ymq}, where, leaving subtleties aside, the relevant ideals can be considerably smaller and their generators far more decoupled.

While not the focus of this Letter, the method also applies to integrals with non-trivial numerators and non-standard (e.g., linearised or eikonal) propagator structures. Indeed, such integrals also admit a Lee--Pomeransky representation. It would be interesting to study, with a large set of examples, cases in which singularities cancel in the presence of numerators. A systematic understanding of such cancellations could have important applications to bootstrap approaches for Standard Model amplitudes.

Finally, we expect the tools introduced in this Letter to converge into an automated \textsc{Mathematica} package.

\phantom{ghost}
\paragraph*{\bf Acknowledgments.}
We would like to thank Miguel Correia, Sebastian Mizera, Erik Panzer, Tiziano Peraro and Bernd Sturmfels for stimulating discussions. We thank Luke Lippstreu, Andrew J. McLeod, and Maria Polackova for allowing us to share functions from the \DiscKosky{} package before release, as well as for many interesting and fruitful conversations. We furthermore thank Franz Herzog and the Institute for Advanced Study for generously lending significant computational resources throughout the duration of this project.
The research of V.C. is funded by the European Research Council (ERC) Synergy Grant MaScAmp 101167287. 
The research of G.C. is supported by the United Kingdom Research and Innovation grant UKRI FLF MR/Y003829/1.
M.G. is supported by the U.S. Department of Energy (DOE) grant No. DE-SC0011941.
\appendix
\section{Subtleties with the Sectors Approach}\label{section:caveats}
\vspace{-0.3cm}
The sector decomposition approach rests on \cref{eq:euler_chi_sector_sum2,eq:euler_chi_sector_sum} holding, which can fail in five known ways. Four of these fall into two families. In the first, a solution set of $\mathcal{J}_\mathcal{S}$ becomes positive-dimensional and, in the second, a critical point escapes to infinity as $\boldsymbol{\nu}\to 0$. Within each family the failure is either generic or confined to a special kinematic slice $\boldsymbol{s}^*$. We label the cases accordingly, as \caveat{1.1} and \caveat{1.2} for positive-dimensional solutions and \caveat{2.1} and \caveat{2.2} for the case of points at infinity. A fifth mechanism, \caveat{3}, arises from sign cancellations between sectors in \cref{eq:inclusion_exclusion}. \caveat{1.1}, \caveat{1.2} and \caveat{2.1} can result in missing genuine singularities, whereas \caveat{2.2} and \caveat{3} can instead generate fictitious ones. All are rare and inexpensive to detect, as explained below. These routines are implemented in the \textsc{Mathematica} package \DiscKosky{} \cite{Crisanti:2026cuf}, with the exception of \texttt{noDegeneracyQ}, which is provided separately in the \zenodorepo{} repository together with all data from this appendix.

As noted in the main text, the examples of \cref{sec:CE} are free of these caveats, the sole exception being a few simple sectors manifesting \caveat{1.2}, whose problematic singularities \SOFIA{} can resolve directly with ease.

\textbf{Type~1.1: General Higher-Dimensional Solutions.}\phantomsection\label{caveat:type1.1}\quad For generic $\boldsymbol{\nu}$, as in \cref{eq:chi_ideal_reg}, the solution set is always zero-dimensional.
At $\boldsymbol{\nu}=0$ this can break: for certain degenerate sector polynomials $\mathcal{G}_\mathcal{S}$, the solutions of $\mathcal{J}_\mathcal{S}$ form positive-dimensional families rather than isolated points. In ideal language, this corresponds to the condition $\dim(\mathcal{J}_\mathcal{S})\neq 0$. This behaviour has also been tied to the ``magic relations'' that break the notion of sectors and cuts \cite{Crisanti:2026rbc}.

\onecolumngrid
\par\vspace{0.4\baselineskip}
\noindent{\color{gray!50}\rule{\textwidth}{0.3pt}}\par
\vspace{0.15\baselineskip}
\begin{center}
  \begin{minipage}[b]{0.27\textwidth}\centering
    \includegraphics[max width=\linewidth]{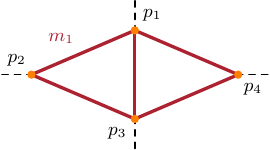}\\[2pt]{\footnotesize (i)}
  \end{minipage}\hfill
  \begin{minipage}[b]{0.27\textwidth}\centering
    \includegraphics[max width=\linewidth]{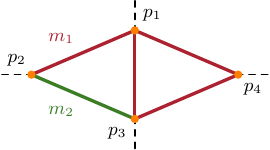}\\[2pt]{\footnotesize (ii)}
  \end{minipage}\hfill
  \begin{minipage}[b]{0.27\textwidth}\centering
    \includegraphics[max width=\linewidth]{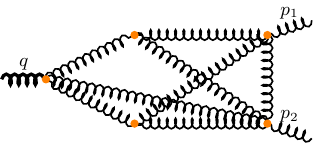}\\[2pt]{\footnotesize (iii)}
  \end{minipage}
  \captionof{figure}{The diagrams discussed in the appendix. All external legs are on-shell ($p_i^2=0$) except $q^2=2 p_1\cdot p_2$, which is off-shell.
  \label{fig:app_diags}
  }
\end{center}
\vspace{-0.9\baselineskip}
\noindent{\color{gray!50}\rule{\textwidth}{0.3pt}}\par
\twocolumngrid
\vspace{0.2\baselineskip}

The diagnostic is therefore to check whether $\dim(\mathcal{J}_\mathcal{S})=0$ on a generic kinematic slice for each sector. A sector with $\dim(\mathcal{J}_\mathcal{S})\neq 0$ signals a failure of \cref{eq:euler_chi_sector_sum}. In practice, $\dim(\mathcal{J}_\mathcal{S})$ is inexpensive to compute, and \DiscKosky{} \cite{Crisanti:2026cuf} evaluates it directly through the function \texttt{CountSectorsUnregulated}, returning \texttt{Indeterminate} if $\dim(\mathcal{J}_\mathcal{S})\neq 0$. An example is the equal-mass ``acnode'' in \cref{fig:app_diags} (i), where the top sector manifests a higher-dimensional solution:
\begin{lstlisting}
vars = {x1, x2, x3, x4, x5};
CountSectorsUnregulated[Gi, vars, vars]
(* Out: Indeterminate *)
\end{lstlisting}

\textbf{Type~1.2: Specialised Higher-Dimensional Solutions.}\phantomsection\label{caveat:type1.2}\quad Even when $\dim(\mathcal{J}_\mathcal{S}(\boldsymbol{s}))=0$ generically, the ideal \textit{dimension} can jump at a special location: $\dim(\mathcal{J}_\mathcal{S}(\boldsymbol{s}^*))\neq 0$. The singularities causing this behaviour must manifest in a very specific (and problematic) form. To see this, let
\begin{equation}
    \mathcal{J}^{(i)}_{\mathcal{S},\text{elim}}\equiv \mathcal{J}_{\mathcal{S}} \cap \mathbb{Q}[\boldsymbol{s}, x_i]\,.
\end{equation}
If $\dim(\mathcal{J}_\mathcal{S}(\boldsymbol{s}^*))\neq 0$, then necessarily for at least one $x_i$\,
\begin{equation}
    \dim(\mathcal{J}^{(i)}_{\mathcal{S},\text{elim}}(\boldsymbol{s}^*))\neq 0\,.
\end{equation}
Since $\mathcal{J}^{(i)}_{\mathcal{S},\text{elim}}$ is generated by a univariate polynomial, it can only be non-zero-dimensional if all coefficients vanish on $\boldsymbol{s}^*$. It follows that
\begin{equation}\label{eq:degenerate_factoring}
    \mathcal{J}^{(i)}_{\mathcal{S},\text{elim}}
    = \left\langle\, l_\text{degen}(\boldsymbol{s})\left
    (\tilde{c}_0^{(i)} + \cdots + \tilde{c}_n^{(i)} x_i^n \right)
    \right\rangle\,,
\end{equation}
where $l_\text{degen}(\boldsymbol{s}^*)=0$. Conservatively, it should be assumed that $l_\text{degen}$ could be a genuine Landau singularity where $\deg(\mathcal{I})$ drops in value. The difficulty is that SP$\mathbb{Q}$R reconstructs Gr\"obner bases over the ring $R = \mathbb{Q}(\boldsymbol{s})[\boldsymbol{x}]\neq\mathbb{Q}[\boldsymbol{s},\boldsymbol{x}]$,
that is, polynomials in the integration variables $\boldsymbol{x}$ whose coefficients are rational functions of the kinematics $\boldsymbol{s}$. Over this ring, every ideal element is defined only up to a kinematic prefactor (fixed by normalising the coefficient of one monomial to $1$ in each generator). The factor $l_\text{degen}$ is therefore absorbed into this arbitrary prefactor, rendering it invisible.

It is thus important to test whether such a singularity can occur. The diagnostic is to recompute the same ideal $\mathcal{J}_\mathcal{S}$ over the ring $\mathbb{Q}(s_2,\ldots)[s_1,\boldsymbol{x}]$, promoting one kinematic variable $s_1$ to a polynomial variable. Elimination over this ring cannot divide out a degenerate singularity that depends on $s_1$, so its presence becomes visible.

In practice, several numerical evaluations on this ring suffice, making this test inexpensive. The algorithm is included in the \zenodorepo{} repository as \texttt{noDegeneracyQ}. If this function returns \texttt{True}, no singularity of the form \cref{eq:degenerate_factoring} can occur. We find that Feynman diagrams violating this condition are rare. The ``deformed acnode'' in \cref{fig:app_diags} (ii)
illustrates this behaviour. One can explicitly verify that
\begin{lstlisting}
noDegeneracyQ[Gii, vars][[1]] (*  Out: False  *)
\end{lstlisting}
flagging that the sector-by-sector elimination is not guaranteed to capture every singularity here. The missed component is $l_\text{degen}=m_1^2-m_2^2$. Indeed, specialising to this kinematic locus results in the example previously considered in \cref{fig:app_diags} (i). Note that singularities manifesting this behaviour seem to be extremely simple, and are captured with ease by \SOFIA{}. This example is provided in more detail in the \zenodorepo{} repository.

\textbf{Type~2.1: Bulk Solutions at Infinity.}\phantomsection\label{caveat:type2.1}\quad The two modes above concern solutions becoming positive-dimensional. A different effect can arise even when $\dim(\mathcal{J}_{\mathcal{S}})=0$: a critical point lying at a finite (bulk) point of $X_{\boldsymbol{s}}$ for generic $\boldsymbol{\nu}$ can be pushed to infinity in the limit $\boldsymbol{\nu}\to0$. Such a point is then absent from $\mathcal{J}_{\mathcal{S}}$, so $\deg(\mathcal{J}_{\mathcal{S}}) < (-1)^{|\mathcal{S}|} \chi(X_{\boldsymbol{s}}^{\mathcal{S}})$  \cite{Bitoun:2017nre}.

The method of checking whether this may occur is very simple. One verifies that \cref{eq:euler_chi_sector_sum} holds before proceeding to the elimination step. Since each $\mathcal{J}_{\mathcal{S}}$ is only known to undercount the Euler characteristic, this equality is sufficient to rule out the issue for all sectors of the diagram. The four-loop form factor given in \cref{fig:app_diags} (iii) realises this phenomenon \cite{Boels:2015yna}.
An explicit computation reveals that
\begin{equation}
    \chi(X_{\boldsymbol{s}})=\deg(\mathcal{I}) = 13\,,\quad \sum_{\mathcal{S}\in \text{sectors}}\deg(\mathcal{J}_{\mathcal{S}}) = 12\,,
\end{equation}
where a critical point attributable to the top sector is manifestly missing in the sector-by-sector approach. Once again, using the package \DiscKosky{} this can be explicitly verified by comparing the outputs of the functions \texttt{CountSectorsUnregulated} and \texttt{CountSectorsRegulated}. Explicitly we have
\begin{lstlisting}
vars = {x1, x2, x3, x4, x5, x6, x7, x8};
idReg = CountSectorsRegulated[Giii,vars,{}];
idSbS = CountSectorsUnregulated[Giii,vars,{}][[1]];
idReg == idSbS (* Out: False *)
\end{lstlisting}

\textbf{Type~2.2: Fictitious Singularities at Infinity.}\phantomsection\label{caveat:type2.2}\quad The sector-by-sector computation can in rare cases generate fictitious singularities. In the elimination step, a locus $\boldsymbol{s}^*$ is flagged whenever a solution of the specialised ideal $\mathcal{J}_\mathcal{S}$ runs to infinity as $\boldsymbol{s}\to\boldsymbol{s}^*$. While this produces genuine singularities in almost all cases, it can happen instead that this behaviour is an artefact of the $\boldsymbol{\nu}=0$ sector specialisation. In this case no regulated critical point is lost, as $\chi$ does not drop at $\boldsymbol{s}^*$:
\begin{equation}
    (-1)^E \chi(X_{\boldsymbol{s}}) = (-1)^E \chi(X_{\boldsymbol{s}^*}) > \sum_{\mathcal{S}\in\text{sectors}}\deg(\mathcal{J}_{\mathcal{S}}(\boldsymbol{s}^*))\,.
\end{equation}
The corresponding factor is thus fictitious. Such examples are not ruled out by the \caveat{2.1} diagnostic, which checks the count $\deg(\mathcal{I}) = \sum_{\mathcal{S}}\deg(\mathcal{J}_\mathcal{S})$ for generic $\boldsymbol{s}$, away from $\boldsymbol{s}^*$.

The diagnostic is again inexpensive. Once the candidate singularities are obtained, each is verified against \cref{eq:euler_chi_drop} at generic $\boldsymbol{\nu}$, and any that does not correspond to a genuine drop is discarded.

An example of this occurring can be found again in the ``deformed acnode'' graph discussed already in the context of \caveat{1.2}. For this diagram, the number of critical points in the top sector drops from $4$ to $3$ on the ``singularity'' given by $l_{\text{spur}}=m_1^2\,s_{12}+m_2^2\,s_{23}+s_{12}\,s_{23}$\,. Nevertheless, a computation with the fully regulated ideal returns no drop:
\begin{lstlisting}
lspur = mm1 s12 + mm2 s23 + s12 s23;
CountSectorsUnregulated[Gii,vars,vars][[1]](* Out:4 *)
CountSectorsUnregulated[Gii,vars,vars,
                "Constraint"->lspur][[1]](* Out:3 *)
CountSectorsRegulated[Gii,vars,{}]      (* Out:30 *)
CountSectorsRegulated[Gii,vars,{},
                  "Constraint"->lspur]  (* Out:30 *)
\end{lstlisting}

\textbf{Type~3: Sign Cancellations Between Sectors.}\phantomsection\label{caveat:type3}\quad The sector-by-sector approach relies on every term $(-1)^{|\mathcal{S}|}\chi(X_{\boldsymbol{s}}^{\mathcal{S}})$ on the right-hand side of \cref{eq:inclusion_exclusion} being non-negative, so that a drop in any single sector is a drop in $\chi(X_{\boldsymbol{s}})$. This is nearly always the case, but does not hold for all Feynman integrals \cite{Bitoun:2017nre}.

When a sector contributes with the opposite sign, a drop in one sector could be compensated by a drop in another, leaving $\chi(X_{\boldsymbol{s}})$ unchanged. The corresponding factor is then spurious. Such terms can be discarded by the same check as \caveat{2.2}, namely verifying each candidate against \cref{eq:euler_chi_drop} at generic $\boldsymbol{\nu}$. We are not aware of a fictitious singularity generated by this mechanism.

\textbf{Diagnostic Pipeline.}\quad Collecting the above checks:
\begin{description}[leftmargin=!]
    \item[\caveat{1.1}] $\dim(\mathcal{J}_{\mathcal{S}})=0$ for every sector $\mathcal{S}$, ruling out general higher-dimensional solutions.
    \item[\caveat{1.2}] \texttt{noDegeneracyQ} returns \texttt{True} for every sector $\mathcal{S}$, ruling out specialised higher-dimensional solutions.
    \item[\caveat{2.1}] $\deg(\mathcal{I}) = \sum_{\mathcal{S}}\deg(\mathcal{J}_\mathcal{S})$ holds explicitly, ruling out bulk solutions at infinity.
    \item[\caveat{2.2}] each candidate singularity is verified against \cref{eq:euler_chi_drop} at generic $\boldsymbol{\nu}$, discarding any fictitious ones.
    \item[\caveat{3}] the \caveat{2.2} check also discards fictitious singularities arising by this mechanism.
\end{description}
With all these tests passed, the method of \cref{sec:method} returns all Landau singularities, barring any as-yet-undiscovered failure mode. For the vast majority of Feynman diagrams these conditions hold, so the scope of the method remains broad. This full diagnostic pipeline is automatically implemented in the demo function \texttt{SPQRLandau}.

Finally, it should be noted that if any of the above criteria are not satisfied, it is still possible to compute $\chi(X_{\boldsymbol{s}})$ directly from the regulated ideal in \cref{eq:chi_ideal_reg}. As argued in the main text, this strategy will in principle work for \textit{all} Feynman integrals. Nevertheless, this will come at some computational cost due to the increased complexity of $\mathcal{I}$ compared to the much simpler set $\mathcal{J}_{\mathcal{S}}$.
  
\vspace{-0.35cm}
\section{Four-Dimensional Kinematics}\label{section:4d}
\vspace{-0.30cm}
In the main text, the six-point singularities are computed with $d$-dimensional external kinematics, imposing no constraints among the Mandelstam invariants besides total momentum conservation. For phenomenological applications one often restricts the external momenta to span at most a four-dimensional subspace (the 't~Hooft--Veltman scheme \cite{tHooft:1972tcz}). This appendix provides an explicit parametrisation.

We consider six massless momenta $p_1,\ldots,p_6$ with $p_i^2=0$ and $\sum_{i=1}^6 p_i=0$. In generic dimension there are nine independent Mandelstam invariants. In four dimensions the Gram matrix $G_{ij}=2\,p_i\cdot p_j$ of $p_1,\ldots,p_5$ must have vanishing determinant, removing one degree of freedom.

We realise this constraint by writing $p_5 = \sum_{i=1}^4 a_i\, p_i$,
which forces $\det G = 0$ identically. The sub-kinematics of $p_1,\ldots,p_4$ is parametrised by six invariants $\{s_{12},s_{23},s_{34},s_{123},s_{234},s_{1234}\}$ with $s_{1234}\equiv s_{56}$, giving
\begin{align}\label{eq:4d_dotproducts}
&
p_i{\cdot}p_{i+1} = \tfrac{s_{i\, i+1}}{2}\,, \,
p_i{\cdot}p_{i+2} = \tfrac{1}{2}(s_{i\,i+1\,i+2}{-}s_{i\,i+1}{-}s_{i+1\,i+2})\,,\notag
\\&
p_1{\cdot}p_4 = \tfrac{1}{2}(s_{1234}{-}s_{123}{-}s_{234}{+}s_{23})\,,
\end{align}
with $1\leq i\leq 3$ in the first relation and $1\leq i\leq 2$ in the second. Imposing $p_5^2=0$ and $p_6^2=0$ (with $p_6=-\sum_{i=1}^5 p_i$) and solving for, e.g., $s_{12}$ and $s_{23}$ yields
\begin{align}\label{eq:4d_gauge}
s_{12} &= \frac{N_{12}}{a_2-a_3}\,,\qquad
s_{23} = \frac{N_{23}}{(a_1{-}a_2)(a_3{-}a_4)}\,,
\end{align}
with the numerators given in the \zenodorepo{} repository.
In this case, the eight free variables are thus $\{s_{34},\, s_{123},\, s_{234},\, s_{1234},\, a_1,\, a_2,\, a_3,\, a_4\}$.
Explicit substitution rules are also provided there.

The substitution \labelcref{eq:4d_gauge} is rational in the external invariants, so specialising the Lee--Pomeransky polynomial $\mathcal{G}$ gives the ratio $\mathcal{G}\big|_{\rm 4d}=\mathcal{G}_{\rm HV}/f(a)$, with $\mathcal{G}_{\rm HV}$ the polynomial numerator and $f(a)$ a purely kinematic denominator. The polynomial $\mathcal{G}_{\rm HV}$ is the 't~Hooft--Veltman polynomial, on which the singularity analysis of \cref{sec:method} can be re-run. The restricted integral therefore reads
\be\label{eq:4d_prefactor}
I_{\boldsymbol{\nu}}\big|_{\rm 4d} \propto f(a)^{d/2} \int_0^\infty \boldsymbol{x}^{\boldsymbol{\nu}}\,\mathcal{G}_{\rm HV}(\boldsymbol{x},\boldsymbol{s})^{-d/2}\,\frac{\d\boldsymbol{x}}{\boldsymbol{x}}\,.
\ee
Because $f(a)$ carries no $\boldsymbol{x}$-dependence, it drops out of the critical-point ideal \labelcref{eq:chi_ideal_reg}, and its zeros therefore need not lower $\chi$. Nonetheless, $f(a)^{d/2}$ generates a genuine branch point of $I_{\boldsymbol{\nu}}$ at $f(a)=0$, so these loci are genuine Landau singularities.

Although specialising kinematic parameters and computing singularities do not commute in general \cite{Fevola:2023kaw}, for diagrams (b, c) of \cref{fig:new_applications} the two procedures agree upon direct computation: (i) restricting the $d$-dimensional singularity list to four-dimensional kinematics, and (ii) re-running the singularity analysis directly on $\mathcal{G}_{\rm HV}$ and adjoining the prefactor singularities $f(a)=0$, produce the same set of singularities.
Apart from these prefactor singularities, every factor arising upon restriction remains genuine in the sense of a $\chi$-drop of \cref{eq:euler_chi_drop}.
  
\bibliography{references1}

\begin{thebibliography}{59}%
\makeatletter
\providecommand \@ifxundefined [1]{%
 \@ifx{#1\undefined}
}%
\providecommand \@ifnum [1]{%
 \ifnum #1\expandafter \@firstoftwo
 \else \expandafter \@secondoftwo
 \fi
}%
\providecommand \@ifx [1]{%
 \ifx #1\expandafter \@firstoftwo
 \else \expandafter \@secondoftwo
 \fi
}%
\providecommand \natexlab [1]{#1}%
\providecommand \enquote  [1]{``#1''}%
\providecommand \bibnamefont  [1]{#1}%
\providecommand \bibfnamefont [1]{#1}%
\providecommand \citenamefont [1]{#1}%
\providecommand \href@noop [0]{\@secondoftwo}%
\providecommand \href [0]{\begingroup \@sanitize@url \@href}%
\providecommand \@href[1]{\@@startlink{#1}\@@href}%
\providecommand \@@href[1]{\endgroup#1\@@endlink}%
\providecommand \@sanitize@url [0]{\catcode `\\12\catcode `\$12\catcode
  `\&12\catcode `\#12\catcode `\^12\catcode `\_12\catcode `\%12\relax}%
\providecommand \@@startlink[1]{}%
\providecommand \@@endlink[0]{}%
\providecommand \url  [0]{\begingroup\@sanitize@url \@url }%
\providecommand \@url [1]{\endgroup\@href {#1}{\urlprefix }}%
\providecommand \urlprefix  [0]{URL }%
\providecommand \Eprint [0]{\href }%
\providecommand \doibase [0]{https://doi.org/}%
\providecommand \selectlanguage [0]{\@gobble}%
\providecommand \bibinfo  [0]{\@secondoftwo}%
\providecommand \bibfield  [0]{\@secondoftwo}%
\providecommand \translation [1]{[#1]}%
\providecommand \BibitemOpen [0]{}%
\providecommand \bibitemStop [0]{}%
\providecommand \bibitemNoStop [0]{.\EOS\space}%
\providecommand \EOS [0]{\spacefactor3000\relax}%
\providecommand \BibitemShut  [1]{\csname bibitem#1\endcsname}%
\let\auto@bib@innerbib\@empty
\bibitem [{\citenamefont {Landau}(1960)}]{Landau:1959fi}%
  \BibitemOpen
  \bibfield  {author} {\bibinfo {author} {\bibfnamefont {L.}~\bibnamefont
  {Landau}},\ }\bibfield  {title} {\bibinfo {title} {{On analytic properties of
  vertex parts in quantum field theory}},\ }\href
  {https://doi.org/10.1016/B978-0-08-010586-4.50103-6} {\bibfield  {journal}
  {\bibinfo  {journal} {Nucl. Phys.}\ }\textbf {\bibinfo {volume} {13}},\
  \bibinfo {pages} {181} (\bibinfo {year} {1960})}\BibitemShut {NoStop}%
\bibitem [{\citenamefont {Nakanishi}(1959)}]{10.1143/PTP.22.128}%
  \BibitemOpen
  \bibfield  {author} {\bibinfo {author} {\bibfnamefont {N.}~\bibnamefont
  {Nakanishi}},\ }\bibfield  {title} {\bibinfo {title} {{Ordinary and Anomalous
  Thresholds in Perturbation Theory}},\ }\href
  {https://doi.org/10.1143/PTP.22.128} {\bibfield  {journal} {\bibinfo
  {journal} {Prog. Theor. Phys.}\ }\textbf {\bibinfo {volume} {22}},\ \bibinfo
  {pages} {128} (\bibinfo {year} {1959})}\BibitemShut {NoStop}%
\bibitem [{\citenamefont {Cutkosky}(1960)}]{Cutkosky:1960sp}%
  \BibitemOpen
  \bibfield  {author} {\bibinfo {author} {\bibfnamefont {R.~E.}\ \bibnamefont
  {Cutkosky}},\ }\bibfield  {title} {\bibinfo {title} {{Singularities and
  discontinuities of Feynman amplitudes}},\ }\href
  {https://doi.org/10.1063/1.1703676} {\bibfield  {journal} {\bibinfo
  {journal} {J. Math. Phys.}\ }\textbf {\bibinfo {volume} {1}},\ \bibinfo
  {pages} {429} (\bibinfo {year} {1960})}\BibitemShut {NoStop}%
\bibitem [{\citenamefont {Collins}(2020)}]{Collins:2020euz}%
  \BibitemOpen
  \bibfield  {author} {\bibinfo {author} {\bibfnamefont {J.}~\bibnamefont
  {Collins}},\ }\bibfield  {title} {\bibinfo {title} {{A new and complete proof
  of the Landau condition for pinch singularities of Feynman graphs and other
  integrals}},\ }\href@noop {} {\bibfield  {journal} {\bibinfo  {journal}
  {arXiv preprint}\ } (\bibinfo {year} {2020})},\ \Eprint
  {https://arxiv.org/abs/2007.04085} {arXiv:2007.04085 [hep-ph]} \BibitemShut
  {NoStop}%
\bibitem [{\citenamefont {Hannesdottir}\ \emph {et~al.}(2022)\citenamefont
  {Hannesdottir}, \citenamefont {McLeod}, \citenamefont {Schwartz},\ and\
  \citenamefont {Vergu}}]{Hannesdottir:2021kpd}%
  \BibitemOpen
  \bibfield  {author} {\bibinfo {author} {\bibfnamefont {H.~S.}\ \bibnamefont
  {Hannesdottir}}, \bibinfo {author} {\bibfnamefont {A.~J.}\ \bibnamefont
  {McLeod}}, \bibinfo {author} {\bibfnamefont {M.~D.}\ \bibnamefont
  {Schwartz}},\ and\ \bibinfo {author} {\bibfnamefont {C.}~\bibnamefont
  {Vergu}},\ }\bibfield  {title} {\bibinfo {title} {{Implications of the Landau
  Equations for Iterated Integrals}},\ }\href
  {https://doi.org/10.1103/PhysRevD.105.L061701} {\bibfield  {journal}
  {\bibinfo  {journal} {Phys. Rev. D}\ }\textbf {\bibinfo {volume} {105}},\
  \bibinfo {pages} {L061701} (\bibinfo {year} {2022})},\ \Eprint
  {https://arxiv.org/abs/2109.09744} {arXiv:2109.09744 [hep-th]} \BibitemShut
  {NoStop}%
\bibitem [{\citenamefont {Dennen}\ \emph {et~al.}(2016)\citenamefont {Dennen},
  \citenamefont {Spradlin},\ and\ \citenamefont {Volovich}}]{Dennen:2015bet}%
  \BibitemOpen
  \bibfield  {author} {\bibinfo {author} {\bibfnamefont {T.}~\bibnamefont
  {Dennen}}, \bibinfo {author} {\bibfnamefont {M.}~\bibnamefont {Spradlin}},\
  and\ \bibinfo {author} {\bibfnamefont {A.}~\bibnamefont {Volovich}},\
  }\bibfield  {title} {\bibinfo {title} {{Landau Singularities and Symbology:
  One- and Two-loop MHV Amplitudes in SYM Theory}},\ }\href
  {https://doi.org/10.1007/JHEP03(2016)069} {\bibfield  {journal} {\bibinfo
  {journal} {JHEP}\ }\textbf {\bibinfo {volume} {03}}\bibfield  {number}
  {\bibinfo  {number} { (3)},\ \bibinfo {pages} {069}},\ }\Eprint
  {https://arxiv.org/abs/1512.07909} {arXiv:1512.07909 [hep-th]} \BibitemShut
  {NoStop}%
\bibitem [{\citenamefont {Dlapa}\ \emph {et~al.}(2023)\citenamefont {Dlapa},
  \citenamefont {Helmer}, \citenamefont {Papathanasiou},\ and\ \citenamefont
  {Tellander}}]{Dlapa:2023cvx}%
  \BibitemOpen
  \bibfield  {author} {\bibinfo {author} {\bibfnamefont {C.}~\bibnamefont
  {Dlapa}}, \bibinfo {author} {\bibfnamefont {M.}~\bibnamefont {Helmer}},
  \bibinfo {author} {\bibfnamefont {G.}~\bibnamefont {Papathanasiou}},\ and\
  \bibinfo {author} {\bibfnamefont {F.}~\bibnamefont {Tellander}},\ }\bibfield
  {title} {\bibinfo {title} {{Symbol Alphabets from the Landau Singular
  Locus}},\ }\href {https://doi.org/10.1007/JHEP10(2023)161} {\bibfield
  {journal} {\bibinfo  {journal} {JHEP}\ }\textbf {\bibinfo {volume}
  {10}}\bibfield  {number} {\bibinfo  {number} { (10)},\ \bibinfo {pages}
  {161}},\ }\Eprint {https://arxiv.org/abs/2304.02629} {arXiv:2304.02629
  [hep-th]} \BibitemShut {NoStop}%
\bibitem [{\citenamefont {Coro}\ \emph {et~al.}(2025)\citenamefont {Coro},
  \citenamefont {Novichkov}, \citenamefont {Page},\ and\ \citenamefont
  {Song}}]{Coro:2025kha}%
  \BibitemOpen
  \bibfield  {author} {\bibinfo {author} {\bibfnamefont {F.}~\bibnamefont
  {Coro}}, \bibinfo {author} {\bibfnamefont {P.~P.}\ \bibnamefont {Novichkov}},
  \bibinfo {author} {\bibfnamefont {B.}~\bibnamefont {Page}},\ and\ \bibinfo
  {author} {\bibfnamefont {Q.}~\bibnamefont {Song}},\ }\href@noop {} {\bibinfo
  {title} {{Feynman Integral Reduction and Landau Singularities}}} (\bibinfo
  {year} {2025}),\ \Eprint {https://arxiv.org/abs/2512.05869} {arXiv:2512.05869
  [hep-th]} \BibitemShut {NoStop}%
\bibitem [{\citenamefont {Correia}\ \emph {et~al.}(2026)\citenamefont
  {Correia}, \citenamefont {Giroux},\ and\ \citenamefont
  {Mizera}}]{Correia:2025yao}%
  \BibitemOpen
  \bibfield  {author} {\bibinfo {author} {\bibfnamefont {M.}~\bibnamefont
  {Correia}}, \bibinfo {author} {\bibfnamefont {M.}~\bibnamefont {Giroux}},\
  and\ \bibinfo {author} {\bibfnamefont {S.}~\bibnamefont {Mizera}},\
  }\bibfield  {title} {\bibinfo {title} {{{\SOFIA}: Singularities of Feynman
  integrals automatized}},\ }\href {https://doi.org/10.1016/j.cpc.2025.109970}
  {\bibfield  {journal} {\bibinfo  {journal} {Comput. Phys. Commun.}\ }\textbf
  {\bibinfo {volume} {320}},\ \bibinfo {pages} {109970} (\bibinfo {year}
  {2026})},\ \Eprint {https://arxiv.org/abs/2503.16601} {arXiv:2503.16601
  [hep-th]} \BibitemShut {NoStop}%
\bibitem [{\citenamefont {Panzer}(2015)}]{Panzer:2014caa}%
  \BibitemOpen
  \bibfield  {author} {\bibinfo {author} {\bibfnamefont {E.}~\bibnamefont
  {Panzer}},\ }\bibfield  {title} {\bibinfo {title} {{Algorithms for the
  symbolic integration of hyperlogarithms with applications to Feynman
  integrals}},\ }\href {https://doi.org/10.1016/j.cpc.2014.10.019} {\bibfield
  {journal} {\bibinfo  {journal} {Comput. Phys. Commun.}\ }\textbf {\bibinfo
  {volume} {188}},\ \bibinfo {pages} {148} (\bibinfo {year} {2015})},\ \Eprint
  {https://arxiv.org/abs/1403.3385} {arXiv:1403.3385 [hep-th]} \BibitemShut
  {NoStop}%
\bibitem [{\citenamefont {Giroux}\ \emph {et~al.}(2026)\citenamefont {Giroux},
  \citenamefont {Mizera},\ and\ \citenamefont {Salvatori}}]{Giroux:2026tgd}%
  \BibitemOpen
  \bibfield  {author} {\bibinfo {author} {\bibfnamefont {M.}~\bibnamefont
  {Giroux}}, \bibinfo {author} {\bibfnamefont {S.}~\bibnamefont {Mizera}},\
  and\ \bibinfo {author} {\bibfnamefont {G.}~\bibnamefont {Salvatori}},\
  }\bibfield  {title} {\bibinfo {title} {{\texttt{SubTropica}}},\ }\href@noop
  {} {\bibfield  {journal} {\bibinfo  {journal} {arXiv preprint}\ } (\bibinfo
  {year} {2026})},\ \Eprint {https://arxiv.org/abs/2604.20954}
  {arXiv:2604.20954 [hep-th]} \BibitemShut {NoStop}%
\bibitem [{\citenamefont {Hannesdottir}\ \emph {et~al.}(2025)\citenamefont
  {Hannesdottir}, \citenamefont {McLeod}, \citenamefont {Schwartz},\ and\
  \citenamefont {Vergu}}]{Hannesdottir:2024hke}%
  \BibitemOpen
  \bibfield  {author} {\bibinfo {author} {\bibfnamefont {H.~S.}\ \bibnamefont
  {Hannesdottir}}, \bibinfo {author} {\bibfnamefont {A.~J.}\ \bibnamefont
  {McLeod}}, \bibinfo {author} {\bibfnamefont {M.~D.}\ \bibnamefont
  {Schwartz}},\ and\ \bibinfo {author} {\bibfnamefont {C.}~\bibnamefont
  {Vergu}},\ }\bibfield  {title} {\bibinfo {title} {{Applications of the Landau
  bootstrap}},\ }\href {https://doi.org/10.1103/PhysRevD.111.085003} {\bibfield
   {journal} {\bibinfo  {journal} {Phys. Rev. D}\ }\textbf {\bibinfo {volume}
  {111}},\ \bibinfo {pages} {085003} (\bibinfo {year} {2025})},\ \Eprint
  {https://arxiv.org/abs/2410.02424} {arXiv:2410.02424 [hep-ph]} \BibitemShut
  {NoStop}%
\bibitem [{\citenamefont {Barrera}\ \emph {et~al.}(2026)\citenamefont
  {Barrera}, \citenamefont {Dersy}, \citenamefont {Husain}, \citenamefont
  {Schwartz},\ and\ \citenamefont {Zhang}}]{Barrera:2025uin}%
  \BibitemOpen
  \bibfield  {author} {\bibinfo {author} {\bibfnamefont {O.}~\bibnamefont
  {Barrera}}, \bibinfo {author} {\bibfnamefont {A.}~\bibnamefont {Dersy}},
  \bibinfo {author} {\bibfnamefont {R.}~\bibnamefont {Husain}}, \bibinfo
  {author} {\bibfnamefont {M.~D.}\ \bibnamefont {Schwartz}},\ and\ \bibinfo
  {author} {\bibfnamefont {X.}~\bibnamefont {Zhang}},\ }\bibfield  {title}
  {\bibinfo {title} {{Analytic regression of Feynman integrals from
  high-precision numerical sampling}},\ }\href
  {https://doi.org/10.1007/JHEP01(2026)014} {\bibfield  {journal} {\bibinfo
  {journal} {JHEP}\ }\textbf {\bibinfo {volume} {01}}\bibfield  {number}
  {\bibinfo  {number} { (1)},\ \bibinfo {pages} {014}},\ }\Eprint
  {https://arxiv.org/abs/2507.17815} {arXiv:2507.17815 [hep-th]} \BibitemShut
  {NoStop}%
\bibitem [{\citenamefont {Bern}\ \emph {et~al.}(1995)\citenamefont {Bern},
  \citenamefont {Dixon}, \citenamefont {Dunbar},\ and\ \citenamefont
  {Kosower}}]{Bern:1994cg}%
  \BibitemOpen
  \bibfield  {author} {\bibinfo {author} {\bibfnamefont {Z.}~\bibnamefont
  {Bern}}, \bibinfo {author} {\bibfnamefont {L.~J.}\ \bibnamefont {Dixon}},
  \bibinfo {author} {\bibfnamefont {D.~C.}\ \bibnamefont {Dunbar}},\ and\
  \bibinfo {author} {\bibfnamefont {D.~A.}\ \bibnamefont {Kosower}},\
  }\bibfield  {title} {\bibinfo {title} {{Fusing gauge theory tree amplitudes
  into loop amplitudes}},\ }\href
  {https://doi.org/10.1016/0550-3213(94)00488-Z} {\bibfield  {journal}
  {\bibinfo  {journal} {Nucl. Phys. B}\ }\textbf {\bibinfo {volume} {435}},\
  \bibinfo {pages} {59} (\bibinfo {year} {1995})},\ \Eprint
  {https://arxiv.org/abs/hep-ph/9409265} {arXiv:hep-ph/9409265} \BibitemShut
  {NoStop}%
\bibitem [{\citenamefont {Caron-Huot}\ \emph {et~al.}(2020)\citenamefont
  {Caron-Huot}, \citenamefont {Dixon}, \citenamefont {Drummond}, \citenamefont
  {Dulat}, \citenamefont {Foster}, \citenamefont {G{\"u}rdo{\u{g}}an},
  \citenamefont {von Hippel}, \citenamefont {McLeod},\ and\ \citenamefont
  {Papathanasiou}}]{Caron-Huot:2020bkp}%
  \BibitemOpen
  \bibfield  {author} {\bibinfo {author} {\bibfnamefont {S.}~\bibnamefont
  {Caron-Huot}}, \bibinfo {author} {\bibfnamefont {L.~J.}\ \bibnamefont
  {Dixon}}, \bibinfo {author} {\bibfnamefont {J.~M.}\ \bibnamefont {Drummond}},
  \bibinfo {author} {\bibfnamefont {F.}~\bibnamefont {Dulat}}, \bibinfo
  {author} {\bibfnamefont {J.}~\bibnamefont {Foster}}, \bibinfo {author}
  {\bibfnamefont {{\"O}.}~\bibnamefont {G{\"u}rdo{\u{g}}an}}, \bibinfo {author}
  {\bibfnamefont {M.}~\bibnamefont {von Hippel}}, \bibinfo {author}
  {\bibfnamefont {A.~J.}\ \bibnamefont {McLeod}},\ and\ \bibinfo {author}
  {\bibfnamefont {G.}~\bibnamefont {Papathanasiou}},\ }\bibfield  {title}
  {\bibinfo {title} {{The Steinmann Cluster Bootstrap for $\mathcal{N}$ = 4
  Super Yang-Mills Amplitudes}},\ }\href {https://doi.org/10.22323/1.376.0003}
  {\bibfield  {journal} {\bibinfo  {journal} {PoS}\ }\textbf {\bibinfo {volume}
  {CORFU2019}},\ \bibinfo {pages} {003} (\bibinfo {year} {2020})},\ \Eprint
  {https://arxiv.org/abs/2005.06735} {arXiv:2005.06735 [hep-th]} \BibitemShut
  {NoStop}%
\bibitem [{\citenamefont {Correia}\ \emph {et~al.}(2021)\citenamefont
  {Correia}, \citenamefont {Sever},\ and\ \citenamefont
  {Zhiboedov}}]{Correia:2020xtr}%
  \BibitemOpen
  \bibfield  {author} {\bibinfo {author} {\bibfnamefont {M.}~\bibnamefont
  {Correia}}, \bibinfo {author} {\bibfnamefont {A.}~\bibnamefont {Sever}},\
  and\ \bibinfo {author} {\bibfnamefont {A.}~\bibnamefont {Zhiboedov}},\
  }\bibfield  {title} {\bibinfo {title} {{An analytical toolkit for the
  S-matrix bootstrap}},\ }\href {https://doi.org/10.1007/JHEP03(2021)013}
  {\bibfield  {journal} {\bibinfo  {journal} {JHEP}\ }\textbf {\bibinfo
  {volume} {03}}\bibfield  {number} {\bibinfo  {number} { (3)},\ \bibinfo
  {pages} {013}},\ }\Eprint {https://arxiv.org/abs/2006.08221}
  {arXiv:2006.08221 [hep-th]} \BibitemShut {NoStop}%
\bibitem [{\citenamefont {Correia}\ \emph {et~al.}(2022)\citenamefont
  {Correia}, \citenamefont {Sever},\ and\ \citenamefont
  {Zhiboedov}}]{Correia:2021etg}%
  \BibitemOpen
  \bibfield  {author} {\bibinfo {author} {\bibfnamefont {M.}~\bibnamefont
  {Correia}}, \bibinfo {author} {\bibfnamefont {A.}~\bibnamefont {Sever}},\
  and\ \bibinfo {author} {\bibfnamefont {A.}~\bibnamefont {Zhiboedov}},\
  }\bibfield  {title} {\bibinfo {title} {{Probing multi-particle unitarity with
  the Landau equations}},\ }\href
  {https://doi.org/10.21468/SciPostPhys.13.3.062} {\bibfield  {journal}
  {\bibinfo  {journal} {SciPost Phys.}\ }\textbf {\bibinfo {volume} {13}},\
  \bibinfo {pages} {062} (\bibinfo {year} {2022})},\ \Eprint
  {https://arxiv.org/abs/2111.12100} {arXiv:2111.12100 [hep-th]} \BibitemShut
  {NoStop}%
\bibitem [{\citenamefont {Carr{\^o}lo}\ \emph {et~al.}(2026)\citenamefont
  {Carr{\^o}lo}, \citenamefont {Chicherin}, \citenamefont {Henn}, \citenamefont
  {Yang},\ and\ \citenamefont {Zhang}}]{Carrolo:2025agz}%
  \BibitemOpen
  \bibfield  {author} {\bibinfo {author} {\bibfnamefont {S.}~\bibnamefont
  {Carr{\^o}lo}}, \bibinfo {author} {\bibfnamefont {D.}~\bibnamefont
  {Chicherin}}, \bibinfo {author} {\bibfnamefont {J.}~\bibnamefont {Henn}},
  \bibinfo {author} {\bibfnamefont {Q.}~\bibnamefont {Yang}},\ and\ \bibinfo
  {author} {\bibfnamefont {Y.}~\bibnamefont {Zhang}},\ }\bibfield  {title}
  {\bibinfo {title} {{Bootstrapping Six-Gluon QCD Amplitudes}},\ }\href
  {https://doi.org/10.1103/k237-8ccq} {\bibfield  {journal} {\bibinfo
  {journal} {Phys. Rev. Lett.}\ }\textbf {\bibinfo {volume} {136}},\ \bibinfo
  {pages} {181602} (\bibinfo {year} {2026})},\ \Eprint
  {https://arxiv.org/abs/2510.20565} {arXiv:2510.20565 [hep-th]} \BibitemShut
  {NoStop}%
\bibitem [{\citenamefont {Mizera}\ and\ \citenamefont
  {Telen}(2022)}]{Mizera:2021icv}%
  \BibitemOpen
  \bibfield  {author} {\bibinfo {author} {\bibfnamefont {S.}~\bibnamefont
  {Mizera}}\ and\ \bibinfo {author} {\bibfnamefont {S.}~\bibnamefont {Telen}},\
  }\bibfield  {title} {\bibinfo {title} {{Landau discriminants}},\ }\href
  {https://doi.org/10.1007/JHEP08(2022)200} {\bibfield  {journal} {\bibinfo
  {journal} {JHEP}\ }\textbf {\bibinfo {volume} {08}}\bibfield  {number}
  {\bibinfo  {number} { (8)},\ \bibinfo {pages} {200}},\ }\Eprint
  {https://arxiv.org/abs/2109.08036} {arXiv:2109.08036 [math-ph]} \BibitemShut
  {NoStop}%
\bibitem [{\citenamefont {Klausen}(2022)}]{Klausen:2021yrt}%
  \BibitemOpen
  \bibfield  {author} {\bibinfo {author} {\bibfnamefont {R.~P.}\ \bibnamefont
  {Klausen}},\ }\bibfield  {title} {\bibinfo {title} {{Kinematic singularities
  of Feynman integrals and principal A-determinants}},\ }\href
  {https://doi.org/10.1007/JHEP02(2022)004} {\bibfield  {journal} {\bibinfo
  {journal} {JHEP}\ }\textbf {\bibinfo {volume} {02}}\bibfield  {number}
  {\bibinfo  {number} { (2)},\ \bibinfo {pages} {004}},\ }\Eprint
  {https://arxiv.org/abs/2109.07584} {arXiv:2109.07584 [hep-th]} \BibitemShut
  {NoStop}%
\bibitem [{\citenamefont {Fevola}\ \emph
  {et~al.}(2024{\natexlab{a}})\citenamefont {Fevola}, \citenamefont {Mizera},\
  and\ \citenamefont {Telen}}]{Fevola:2023kaw}%
  \BibitemOpen
  \bibfield  {author} {\bibinfo {author} {\bibfnamefont {C.}~\bibnamefont
  {Fevola}}, \bibinfo {author} {\bibfnamefont {S.}~\bibnamefont {Mizera}},\
  and\ \bibinfo {author} {\bibfnamefont {S.}~\bibnamefont {Telen}},\ }\bibfield
   {title} {\bibinfo {title} {{Landau Singularities Revisited: Computational
  Algebraic Geometry for Feynman Integrals}},\ }\href
  {https://doi.org/10.1103/PhysRevLett.132.101601} {\bibfield  {journal}
  {\bibinfo  {journal} {Phys. Rev. Lett.}\ }\textbf {\bibinfo {volume} {132}},\
  \bibinfo {pages} {101601} (\bibinfo {year} {2024}{\natexlab{a}})},\ \Eprint
  {https://arxiv.org/abs/2311.14669} {arXiv:2311.14669 [hep-th]} \BibitemShut
  {NoStop}%
\bibitem [{\citenamefont {Fevola}\ \emph
  {et~al.}(2024{\natexlab{b}})\citenamefont {Fevola}, \citenamefont {Mizera},\
  and\ \citenamefont {Telen}}]{Fevola:2023fzn}%
  \BibitemOpen
  \bibfield  {author} {\bibinfo {author} {\bibfnamefont {C.}~\bibnamefont
  {Fevola}}, \bibinfo {author} {\bibfnamefont {S.}~\bibnamefont {Mizera}},\
  and\ \bibinfo {author} {\bibfnamefont {S.}~\bibnamefont {Telen}},\ }\bibfield
   {title} {\bibinfo {title} {{Principal Landau determinants}},\ }\href
  {https://doi.org/10.1016/j.cpc.2024.109278} {\bibfield  {journal} {\bibinfo
  {journal} {Comput. Phys. Commun.}\ }\textbf {\bibinfo {volume} {303}},\
  \bibinfo {pages} {109278} (\bibinfo {year} {2024}{\natexlab{b}})},\ \Eprint
  {https://arxiv.org/abs/2311.16219} {arXiv:2311.16219 [math-ph]} \BibitemShut
  {NoStop}%
\bibitem [{\citenamefont {Helmer}\ \emph {et~al.}(2024)\citenamefont {Helmer},
  \citenamefont {Papathanasiou},\ and\ \citenamefont
  {Tellander}}]{Helmer:2024wax}%
  \BibitemOpen
  \bibfield  {author} {\bibinfo {author} {\bibfnamefont {M.}~\bibnamefont
  {Helmer}}, \bibinfo {author} {\bibfnamefont {G.}~\bibnamefont
  {Papathanasiou}},\ and\ \bibinfo {author} {\bibfnamefont {F.}~\bibnamefont
  {Tellander}},\ }\bibfield  {title} {\bibinfo {title} {{Landau Singularities
  from Whitney Stratifications}},\ }\href@noop {} {\bibfield  {journal}
  {\bibinfo  {journal} {arXiv preprint}\ } (\bibinfo {year} {2024})},\ \Eprint
  {https://arxiv.org/abs/2402.14787} {arXiv:2402.14787 [hep-th]} \BibitemShut
  {NoStop}%
\bibitem [{\citenamefont {Helmer}\ and\ \citenamefont
  {Tellander}(2026)}]{Helmer:2025yuf}%
  \BibitemOpen
  \bibfield  {author} {\bibinfo {author} {\bibfnamefont {M.}~\bibnamefont
  {Helmer}}\ and\ \bibinfo {author} {\bibfnamefont {F.}~\bibnamefont
  {Tellander}},\ }\bibfield  {title} {\bibinfo {title} {{Spectral Decomposition
  of Euler-Mellin Integrals}},\ }\href
  {https://doi.org/10.1007/s00220-026-05692-4} {\bibfield  {journal} {\bibinfo
  {journal} {Commun. Math. Phys.}\ }\textbf {\bibinfo {volume} {407}},\
  \bibinfo {pages} {223} (\bibinfo {year} {2026})},\ \Eprint
  {https://arxiv.org/abs/2505.12458} {arXiv:2505.12458 [math-ph]} \BibitemShut
  {NoStop}%
\bibitem [{\citenamefont {Helmer}\ and\ \citenamefont
  {Tellander}(2025)}]{Helmer:2025ljj}%
  \BibitemOpen
  \bibfield  {author} {\bibinfo {author} {\bibfnamefont {M.}~\bibnamefont
  {Helmer}}\ and\ \bibinfo {author} {\bibfnamefont {F.}~\bibnamefont
  {Tellander}},\ }\bibfield  {title} {\bibinfo {title} {{Geometric
  singularities of Feynman integrals}},\ }\href
  {https://doi.org/10.1103/l234-l557} {\bibfield  {journal} {\bibinfo
  {journal} {Phys. Rev. D}\ }\textbf {\bibinfo {volume} {112}},\ \bibinfo
  {pages} {065019} (\bibinfo {year} {2025})},\ \Eprint
  {https://arxiv.org/abs/2506.05042} {arXiv:2506.05042 [hep-th]} \BibitemShut
  {NoStop}%
\bibitem [{\citenamefont {Telen}\ and\ \citenamefont
  {Wiesmann}(2024)}]{Telen:2024sep}%
  \BibitemOpen
  \bibfield  {author} {\bibinfo {author} {\bibfnamefont {S.}~\bibnamefont
  {Telen}}\ and\ \bibinfo {author} {\bibfnamefont {M.}~\bibnamefont
  {Wiesmann}},\ }\bibfield  {title} {\bibinfo {title} {{Euler Stratifications
  of Hypersurface Families}},\ }\href@noop {} {\bibfield  {journal} {\bibinfo
  {journal} {arXiv preprint}\ } (\bibinfo {year} {2024})},\ \Eprint
  {https://arxiv.org/abs/2407.18176} {arXiv:2407.18176 [math.AG]} \BibitemShut
  {NoStop}%
\bibitem [{\citenamefont {Caron-Huot}\ \emph {et~al.}(2025)\citenamefont
  {Caron-Huot}, \citenamefont {Correia},\ and\ \citenamefont
  {Giroux}}]{Caron-Huot:2024brh}%
  \BibitemOpen
  \bibfield  {author} {\bibinfo {author} {\bibfnamefont {S.}~\bibnamefont
  {Caron-Huot}}, \bibinfo {author} {\bibfnamefont {M.}~\bibnamefont
  {Correia}},\ and\ \bibinfo {author} {\bibfnamefont {M.}~\bibnamefont
  {Giroux}},\ }\bibfield  {title} {\bibinfo {title} {{Recursive Landau
  Analysis}},\ }\href {https://doi.org/10.1103/8rwk-bnph} {\bibfield  {journal}
  {\bibinfo  {journal} {Phys. Rev. Lett.}\ }\textbf {\bibinfo {volume} {135}},\
  \bibinfo {pages} {131603} (\bibinfo {year} {2025})},\ \Eprint
  {https://arxiv.org/abs/2406.05241} {arXiv:2406.05241 [hep-th]} \BibitemShut
  {NoStop}%
\bibitem [{\citenamefont {Lee}\ and\ \citenamefont
  {Pomeransky}(2013)}]{Lee:2013hzt}%
  \BibitemOpen
  \bibfield  {author} {\bibinfo {author} {\bibfnamefont {R.~N.}\ \bibnamefont
  {Lee}}\ and\ \bibinfo {author} {\bibfnamefont {A.~A.}\ \bibnamefont
  {Pomeransky}},\ }\bibfield  {title} {\bibinfo {title} {{Critical points and
  number of master integrals}},\ }\href
  {https://doi.org/10.1007/JHEP11(2013)165} {\bibfield  {journal} {\bibinfo
  {journal} {JHEP}\ }\textbf {\bibinfo {volume} {11}}\bibfield  {number}
  {\bibinfo  {number} { (11)},\ \bibinfo {pages} {165}},\ }\Eprint
  {https://arxiv.org/abs/1308.6676} {arXiv:1308.6676 [hep-ph]} \BibitemShut
  {NoStop}%
\bibitem [{\citenamefont {Frellesvig}\ \emph
  {et~al.}(2019{\natexlab{a}})\citenamefont {Frellesvig}, \citenamefont
  {Gasparotto}, \citenamefont {Laporta}, \citenamefont {Mandal}, \citenamefont
  {Mastrolia}, \citenamefont {Mattiazzi},\ and\ \citenamefont
  {Mizera}}]{Frellesvig:2019kgj}%
  \BibitemOpen
  \bibfield  {author} {\bibinfo {author} {\bibfnamefont {H.}~\bibnamefont
  {Frellesvig}}, \bibinfo {author} {\bibfnamefont {F.}~\bibnamefont
  {Gasparotto}}, \bibinfo {author} {\bibfnamefont {S.}~\bibnamefont {Laporta}},
  \bibinfo {author} {\bibfnamefont {M.~K.}\ \bibnamefont {Mandal}}, \bibinfo
  {author} {\bibfnamefont {P.}~\bibnamefont {Mastrolia}}, \bibinfo {author}
  {\bibfnamefont {L.}~\bibnamefont {Mattiazzi}},\ and\ \bibinfo {author}
  {\bibfnamefont {S.}~\bibnamefont {Mizera}},\ }\bibfield  {title} {\bibinfo
  {title} {{Decomposition of Feynman Integrals on the Maximal Cut by
  Intersection Numbers}},\ }\href {https://doi.org/10.1007/JHEP05(2019)153}
  {\bibfield  {journal} {\bibinfo  {journal} {JHEP}\ }\textbf {\bibinfo
  {volume} {05}}\bibfield  {number} {\bibinfo  {number} { (5)},\ \bibinfo
  {pages} {153}},\ }\Eprint {https://arxiv.org/abs/1901.11510}
  {arXiv:1901.11510 [hep-ph]} \BibitemShut {NoStop}%
\bibitem [{\citenamefont {Matsubara-Heo}(2025)}]{Matsubara-Heo:2025lrq}%
  \BibitemOpen
  \bibfield  {author} {\bibinfo {author} {\bibfnamefont {S.-J.}\ \bibnamefont
  {Matsubara-Heo}},\ }\href@noop {} {\bibinfo {title} {{Hypergeometric
  Discriminants}}} (\bibinfo {year} {2025}),\ \Eprint
  {https://arxiv.org/abs/2505.13163} {arXiv:2505.13163 [math.AG]} \BibitemShut
  {NoStop}%
\bibitem [{\citenamefont {Chestnov}\ and\ \citenamefont
  {Crisanti}(2025)}]{Chestnov:2025svg}%
  \BibitemOpen
  \bibfield  {author} {\bibinfo {author} {\bibfnamefont {V.}~\bibnamefont
  {Chestnov}}\ and\ \bibinfo {author} {\bibfnamefont {G.}~\bibnamefont
  {Crisanti}},\ }\href@noop {} {\bibinfo {title} {{Sampling Polynomial Rational
  Remainders with SP$\mathbb{Q}$R: A new Package for Polynomial Division and
  Elimination}}} (\bibinfo {year} {2025}),\ \Eprint
  {https://arxiv.org/abs/2511.14875} {arXiv:2511.14875 [hep-th]} \BibitemShut
  {NoStop}%
\bibitem [{\citenamefont {Peraro}(2016)}]{Peraro:2016wsq}%
  \BibitemOpen
  \bibfield  {author} {\bibinfo {author} {\bibfnamefont {T.}~\bibnamefont
  {Peraro}},\ }\bibfield  {title} {\bibinfo {title} {{Scattering amplitudes
  over finite fields and multivariate functional reconstruction}},\ }\href
  {https://doi.org/10.1007/JHEP12(2016)030} {\bibfield  {journal} {\bibinfo
  {journal} {JHEP}\ }\textbf {\bibinfo {volume} {12}}\bibfield  {number}
  {\bibinfo  {number} { (12)},\ \bibinfo {pages} {030}},\ }\Eprint
  {https://arxiv.org/abs/1608.01902} {arXiv:1608.01902 [hep-ph]} \BibitemShut
  {NoStop}%
\bibitem [{\citenamefont {Peraro}(2019)}]{Peraro:2019svx}%
  \BibitemOpen
  \bibfield  {author} {\bibinfo {author} {\bibfnamefont {T.}~\bibnamefont
  {Peraro}},\ }\bibfield  {title} {\bibinfo {title} {{$\text{FiniteFlow}$:
  multivariate functional reconstruction using finite fields and dataflow
  graphs}},\ }\href {https://doi.org/10.1007/JHEP07(2019)031} {\bibfield
  {journal} {\bibinfo  {journal} {JHEP}\ }\textbf {\bibinfo {volume}
  {07}}\bibfield  {number} {\bibinfo  {number} { (7)},\ \bibinfo {pages}
  {031}},\ }\Eprint {https://arxiv.org/abs/1905.08019} {arXiv:1905.08019
  [hep-ph]} \BibitemShut {NoStop}%
\bibitem [{\citenamefont {Baikov}(1996)}]{Baikov:1996cd}%
  \BibitemOpen
  \bibfield  {author} {\bibinfo {author} {\bibfnamefont {P.~A.}\ \bibnamefont
  {Baikov}},\ }\bibfield  {title} {\bibinfo {title} {{Explicit solutions of n
  loop vacuum integral recurrence relations}},\ }\href@noop {} {\bibfield
  {journal} {\bibinfo  {journal} {arXiv preprint}\ } (\bibinfo {year}
  {1996})},\ \Eprint {https://arxiv.org/abs/hep-ph/9604254}
  {arXiv:hep-ph/9604254} \BibitemShut {NoStop}%
\bibitem [{\citenamefont {Baikov}(2006)}]{Baikov:2005nv}%
  \BibitemOpen
  \bibfield  {author} {\bibinfo {author} {\bibfnamefont {P.~A.}\ \bibnamefont
  {Baikov}},\ }\bibfield  {title} {\bibinfo {title} {{A Practical criterion of
  irreducibility of multi-loop Feynman integrals}},\ }\href
  {https://doi.org/10.1016/j.physletb.2006.01.052} {\bibfield  {journal}
  {\bibinfo  {journal} {Phys. Lett. B}\ }\textbf {\bibinfo {volume} {634}},\
  \bibinfo {pages} {325} (\bibinfo {year} {2006})},\ \Eprint
  {https://arxiv.org/abs/hep-ph/0507053} {arXiv:hep-ph/0507053} \BibitemShut
  {NoStop}%
\bibitem [{\citenamefont {Frellesvig}(2025)}]{Frellesvig:2024ymq}%
  \BibitemOpen
  \bibfield  {author} {\bibinfo {author} {\bibfnamefont {H.}~\bibnamefont
  {Frellesvig}},\ }\bibfield  {title} {\bibinfo {title} {{The loop-by-loop
  Baikov representation {\textemdash} Strategies and implementation}},\ }\href
  {https://doi.org/10.1007/JHEP04(2025)111} {\bibfield  {journal} {\bibinfo
  {journal} {JHEP}\ }\textbf {\bibinfo {volume} {04}}\bibfield  {number}
  {\bibinfo  {number} { (04)},\ \bibinfo {pages} {111}},\ }\Eprint
  {https://arxiv.org/abs/2412.01804} {arXiv:2412.01804 [hep-th]} \BibitemShut
  {NoStop}%
\bibitem [{\citenamefont {Matsubara-Heo}\ \emph {et~al.}(2023)\citenamefont
  {Matsubara-Heo}, \citenamefont {Mizera},\ and\ \citenamefont
  {Telen}}]{Matsubara-Heo:2023ylc}%
  \BibitemOpen
  \bibfield  {author} {\bibinfo {author} {\bibfnamefont {S.-J.}\ \bibnamefont
  {Matsubara-Heo}}, \bibinfo {author} {\bibfnamefont {S.}~\bibnamefont
  {Mizera}},\ and\ \bibinfo {author} {\bibfnamefont {S.}~\bibnamefont
  {Telen}},\ }\bibfield  {title} {\bibinfo {title} {{Four lectures on Euler
  integrals}},\ }\href {https://doi.org/10.21468/SciPostPhysLectNotes.75}
  {\bibfield  {journal} {\bibinfo  {journal} {SciPost Phys. Lect. Notes}\
  }\textbf {\bibinfo {volume} {75}},\ \bibinfo {pages} {1} (\bibinfo {year}
  {2023})},\ \Eprint {https://arxiv.org/abs/2306.13578} {arXiv:2306.13578
  [math-ph]} \BibitemShut {NoStop}%
\bibitem [{\citenamefont {Huh}(2013)}]{Huh:2013}%
  \BibitemOpen
  \bibfield  {author} {\bibinfo {author} {\bibfnamefont {J.}~\bibnamefont
  {Huh}},\ }\bibfield  {title} {\bibinfo {title} {The maximum likelihood degree
  of a very affine variety},\ }\href
  {https://doi.org/10.1112/S0010437X13007057} {\bibfield  {journal} {\bibinfo
  {journal} {Compositio Mathematica}\ }\textbf {\bibinfo {volume} {149}},\
  \bibinfo {pages} {1245} (\bibinfo {year} {2013})},\ \Eprint
  {https://arxiv.org/abs/1207.0553} {arXiv:1207.0553 [math.AG]} \BibitemShut
  {NoStop}%
\bibitem [{\citenamefont {Bitoun}\ \emph {et~al.}(2019)\citenamefont {Bitoun},
  \citenamefont {Bogner}, \citenamefont {Klausen},\ and\ \citenamefont
  {Panzer}}]{Bitoun:2017nre}%
  \BibitemOpen
  \bibfield  {author} {\bibinfo {author} {\bibfnamefont {T.}~\bibnamefont
  {Bitoun}}, \bibinfo {author} {\bibfnamefont {C.}~\bibnamefont {Bogner}},
  \bibinfo {author} {\bibfnamefont {R.~P.}\ \bibnamefont {Klausen}},\ and\
  \bibinfo {author} {\bibfnamefont {E.}~\bibnamefont {Panzer}},\ }\bibfield
  {title} {\bibinfo {title} {{Feynman integral relations from parametric
  annihilators}},\ }\href {https://doi.org/10.1007/s11005-018-1114-8}
  {\bibfield  {journal} {\bibinfo  {journal} {Lett. Math. Phys.}\ }\textbf
  {\bibinfo {volume} {109}},\ \bibinfo {pages} {497} (\bibinfo {year}
  {2019})},\ \Eprint {https://arxiv.org/abs/1712.09215} {arXiv:1712.09215
  [hep-th]} \BibitemShut {NoStop}%
\bibitem [{\citenamefont {Frellesvig}\ \emph
  {et~al.}(2019{\natexlab{b}})\citenamefont {Frellesvig}, \citenamefont
  {Gasparotto}, \citenamefont {Mandal}, \citenamefont {Mastrolia},
  \citenamefont {Mattiazzi},\ and\ \citenamefont
  {Mizera}}]{Frellesvig:2019uqt}%
  \BibitemOpen
  \bibfield  {author} {\bibinfo {author} {\bibfnamefont {H.}~\bibnamefont
  {Frellesvig}}, \bibinfo {author} {\bibfnamefont {F.}~\bibnamefont
  {Gasparotto}}, \bibinfo {author} {\bibfnamefont {M.~K.}\ \bibnamefont
  {Mandal}}, \bibinfo {author} {\bibfnamefont {P.}~\bibnamefont {Mastrolia}},
  \bibinfo {author} {\bibfnamefont {L.}~\bibnamefont {Mattiazzi}},\ and\
  \bibinfo {author} {\bibfnamefont {S.}~\bibnamefont {Mizera}},\ }\bibfield
  {title} {\bibinfo {title} {{Vector Space of Feynman Integrals and
  Multivariate Intersection Numbers}},\ }\href
  {https://doi.org/10.1103/PhysRevLett.123.201602} {\bibfield  {journal}
  {\bibinfo  {journal} {Phys. Rev. Lett.}\ }\textbf {\bibinfo {volume} {123}},\
  \bibinfo {pages} {201602} (\bibinfo {year} {2019}{\natexlab{b}})},\ \Eprint
  {https://arxiv.org/abs/1907.02000} {arXiv:1907.02000 [hep-th]} \BibitemShut
  {NoStop}%
\bibitem [{\citenamefont {Franecki}\ and\ \citenamefont
  {Kapranov}(1999)}]{Franecki:1999gcm}%
  \BibitemOpen
  \bibfield  {author} {\bibinfo {author} {\bibfnamefont {J.}~\bibnamefont
  {Franecki}}\ and\ \bibinfo {author} {\bibfnamefont {M.}~\bibnamefont
  {Kapranov}},\ }\bibfield  {title} {\bibinfo {title} {{The Gauss map and a
  noncompact Riemann-Roch formula for constructible sheaves on semiabelian
  varieties}},\ }\href@noop {} {\bibfield  {journal} {\bibinfo  {journal}
  {arXiv preprint}\ } (\bibinfo {year} {1999})},\ \Eprint
  {https://arxiv.org/abs/math/9909088} {arXiv:math/9909088 [math.AG]}
  \BibitemShut {NoStop}%
\bibitem [{\citenamefont {Laporta}\ and\ \citenamefont
  {Remiddi}(2005)}]{Laporta:2004rb}%
  \BibitemOpen
  \bibfield  {author} {\bibinfo {author} {\bibfnamefont {S.}~\bibnamefont
  {Laporta}}\ and\ \bibinfo {author} {\bibfnamefont {E.}~\bibnamefont
  {Remiddi}},\ }\bibfield  {title} {\bibinfo {title} {{Analytic treatment of
  the two loop equal mass sunrise graph}},\ }\href
  {https://doi.org/10.1016/j.nuclphysb.2004.10.044} {\bibfield  {journal}
  {\bibinfo  {journal} {Nucl. Phys. B}\ }\textbf {\bibinfo {volume} {704}},\
  \bibinfo {pages} {349} (\bibinfo {year} {2005})},\ \Eprint
  {https://arxiv.org/abs/hep-ph/0406160} {arXiv:hep-ph/0406160 [hep-ph]}
  \BibitemShut {NoStop}%
\bibitem [{\citenamefont {Fevola}\ \emph {et~al.}(2023)\citenamefont {Fevola},
  \citenamefont {Mizera},\ and\ \citenamefont {Telen}}]{PLDdata}%
  \BibitemOpen
  \bibfield  {author} {\bibinfo {author} {\bibfnamefont {C.}~\bibnamefont
  {Fevola}}, \bibinfo {author} {\bibfnamefont {S.}~\bibnamefont {Mizera}},\
  and\ \bibinfo {author} {\bibfnamefont {S.}~\bibnamefont {Telen}},\
  }\href@noop {} {\bibinfo {title} {{Principal Landau Determinants:
  accompanying data}}},\ \bibinfo {howpublished} {MathRepo, Max Planck
  Institute for Mathematics in the Sciences} (\bibinfo {year} {2023}),\
  \bibinfo {note} {landau singularity data for the envelope (a) and non-planar
  double-pentagon (b,c) diagrams: \\
  \url{https://mathrepo.mis.mpg.de/_downloads/88c0f5a43ba3daaccdd9c05510d7ee4d/env_equal_equal.txt},
  \\
  \url{https://mathrepo.mis.mpg.de/_downloads/fc7ed2a5b38ff9bac31711c9231a4bcc/npl-dpent2_zero_zero.txt},
  \\
  \url{https://mathrepo.mis.mpg.de/_downloads/e1b3d0d81f748f9d22a2347992951d4c/npl-dpent_zero_zero.txt}}\BibitemShut
  {NoStop}%
\bibitem [{Note1()}]{Note1}%
  \BibitemOpen
  \bibinfo {note} {We furthermore independently verified the drop in the number
  of master integrals on these singularities using IBP identities~\cite
  {Tkachov:1981wb,Chetyrkin:1981qh} on fully numerical slices, generated with
  the private \protect \textsc {Mathematica} package~\protect \textsc
  {FFIntRed} by Tiziano Peraro and solved by a variant of the Laporta
  algorithm~\cite {Laporta:2001dd} based on the finite-field arithmetic and
  functional reconstruction framework of~\protect \textsc {FiniteFlow}~\cite
  {Peraro:2019svx}.}\BibitemShut {Stop}%
\bibitem [{\citenamefont {'t~Hooft}\ and\ \citenamefont
  {Veltman}(1972)}]{tHooft:1972tcz}%
  \BibitemOpen
  \bibfield  {author} {\bibinfo {author} {\bibfnamefont {G.}~\bibnamefont
  {'t~Hooft}}\ and\ \bibinfo {author} {\bibfnamefont {M.~J.~G.}\ \bibnamefont
  {Veltman}},\ }\bibfield  {title} {\bibinfo {title} {{Regularization and
  Renormalization of Gauge Fields}},\ }\href
  {https://doi.org/10.1016/0550-3213(72)90279-9} {\bibfield  {journal}
  {\bibinfo  {journal} {Nucl. Phys. B}\ }\textbf {\bibinfo {volume} {44}},\
  \bibinfo {pages} {189} (\bibinfo {year} {1972})}\BibitemShut {NoStop}%
\bibitem [{Note2()}]{Note2}%
  \BibitemOpen
  \bibinfo {note} {We also cross-checked the corresponding Euler
  characteristics for each sector using IBP identities, following the strategy
  of~\cite {Henn:2024ngj}.}\BibitemShut {Stop}%
\bibitem [{rep(2025)}]{repoEffortless}%
  \BibitemOpen
  \href@noop {} {}\bibinfo {howpublished}
  {\href{https://github.com/antonela-matijasic/Effortless/tree/main}{Effortless
  GitHub repository}} (\bibinfo {year} {2025}),\ \bibinfo {note} {{G}itHub
  repository, accessed: 2025-02-24}\BibitemShut {NoStop}%
\bibitem [{\citenamefont {Abreu}\ \emph {et~al.}(2025)\citenamefont {Abreu},
  \citenamefont {Monni}, \citenamefont {Page},\ and\ \citenamefont
  {Usovitsch}}]{Abreu:2024fei}%
  \BibitemOpen
  \bibfield  {author} {\bibinfo {author} {\bibfnamefont {S.}~\bibnamefont
  {Abreu}}, \bibinfo {author} {\bibfnamefont {P.~F.}\ \bibnamefont {Monni}},
  \bibinfo {author} {\bibfnamefont {B.}~\bibnamefont {Page}},\ and\ \bibinfo
  {author} {\bibfnamefont {J.}~\bibnamefont {Usovitsch}},\ }\bibfield  {title}
  {\bibinfo {title} {{Planar six-point Feynman integrals for four-dimensional
  gauge theories}},\ }\href {https://doi.org/10.1007/JHEP06(2025)112}
  {\bibfield  {journal} {\bibinfo  {journal} {JHEP}\ }\textbf {\bibinfo
  {volume} {06}}\bibfield  {number} {\bibinfo  {number} { (112)},\ \bibinfo
  {pages} {112}},\ }\Eprint {https://arxiv.org/abs/2412.19884}
  {arXiv:2412.19884 [hep-ph]} \BibitemShut {NoStop}%
\bibitem [{\citenamefont {Henn}\ \emph {et~al.}(2025)\citenamefont {Henn},
  \citenamefont {Matija{\v{s}}i{\'c}}, \citenamefont {Miczajka}, \citenamefont
  {Peraro}, \citenamefont {Xu},\ and\ \citenamefont {Zhang}}]{Henn:2025xrc}%
  \BibitemOpen
  \bibfield  {author} {\bibinfo {author} {\bibfnamefont {J.}~\bibnamefont
  {Henn}}, \bibinfo {author} {\bibfnamefont {A.}~\bibnamefont
  {Matija{\v{s}}i{\'c}}}, \bibinfo {author} {\bibfnamefont {J.}~\bibnamefont
  {Miczajka}}, \bibinfo {author} {\bibfnamefont {T.}~\bibnamefont {Peraro}},
  \bibinfo {author} {\bibfnamefont {Y.}~\bibnamefont {Xu}},\ and\ \bibinfo
  {author} {\bibfnamefont {Y.}~\bibnamefont {Zhang}},\ }\bibfield  {title}
  {\bibinfo {title} {{Complete Function Space for Planar Two-Loop Six-Particle
  Scattering Amplitudes}},\ }\href {https://doi.org/10.1103/zhzd-tj9p}
  {\bibfield  {journal} {\bibinfo  {journal} {Phys. Rev. Lett.}\ }\textbf
  {\bibinfo {volume} {135}},\ \bibinfo {pages} {031601} (\bibinfo {year}
  {2025})},\ \Eprint {https://arxiv.org/abs/2501.01847} {arXiv:2501.01847
  [hep-ph]} \BibitemShut {NoStop}%
\bibitem [{\citenamefont {Liu}\ \emph {et~al.}(2026)\citenamefont {Liu},
  \citenamefont {Matija{\v{s}}i{\'c}}, \citenamefont {Peraro}, \citenamefont
  {Xu}, \citenamefont {Yang},\ and\ \citenamefont {Zhang}}]{Liu:2026hdp}%
  \BibitemOpen
  \bibfield  {author} {\bibinfo {author} {\bibfnamefont {Y.}~\bibnamefont
  {Liu}}, \bibinfo {author} {\bibfnamefont {A.}~\bibnamefont
  {Matija{\v{s}}i{\'c}}}, \bibinfo {author} {\bibfnamefont {T.}~\bibnamefont
  {Peraro}}, \bibinfo {author} {\bibfnamefont {Y.}~\bibnamefont {Xu}}, \bibinfo
  {author} {\bibfnamefont {Z.}~\bibnamefont {Yang}},\ and\ \bibinfo {author}
  {\bibfnamefont {Y.}~\bibnamefont {Zhang}},\ }\bibfield  {title} {\bibinfo
  {title} {{Two-loop Six-point Planar Massless Feynman Integrals to Higher
  $\varepsilon$ Orders}},\ }\href@noop {} {\bibfield  {journal} {\bibinfo
  {journal} {arXiv preprint}\ } (\bibinfo {year} {2026})},\ \Eprint
  {https://arxiv.org/abs/2603.16831} {arXiv:2603.16831 [hep-ph]} \BibitemShut
  {NoStop}%
\bibitem [{\citenamefont {Berthomieu}\ \emph {et~al.}(2021)\citenamefont
  {Berthomieu}, \citenamefont {Eder},\ and\ \citenamefont {{Safey El
  Din}}}]{msolve}%
  \BibitemOpen
  \bibfield  {author} {\bibinfo {author} {\bibfnamefont {J.}~\bibnamefont
  {Berthomieu}}, \bibinfo {author} {\bibfnamefont {C.}~\bibnamefont {Eder}},\
  and\ \bibinfo {author} {\bibfnamefont {M.}~\bibnamefont {{Safey El Din}}},\
  }\bibfield  {title} {\bibinfo {title} {{\texttt{msolve}: A Library for
  Solving Polynomial Systems}},\ }in\ \href
  {https://doi.org/10.1145/3452143.3465545} {\emph {\bibinfo {booktitle} {{2021
  International Symposium on Symbolic and Algebraic Computation}}}},\ \bibinfo
  {series and number} {46th International Symposium on Symbolic and Algebraic
  Computation}\ (\bibinfo  {publisher} {{ACM}},\ \bibinfo {address} {Saint
  Petersburg, Russia},\ \bibinfo {year} {2021})\ pp.\ \bibinfo {pages}
  {51--58}\BibitemShut {NoStop}%
\bibitem [{\citenamefont {Frellesvig}\ and\ \citenamefont
  {Papadopoulos}(2017)}]{Frellesvig:2017aai}%
  \BibitemOpen
  \bibfield  {author} {\bibinfo {author} {\bibfnamefont {H.}~\bibnamefont
  {Frellesvig}}\ and\ \bibinfo {author} {\bibfnamefont {C.~G.}\ \bibnamefont
  {Papadopoulos}},\ }\bibfield  {title} {\bibinfo {title} {{Cuts of Feynman
  Integrals in Baikov representation}},\ }\href
  {https://doi.org/10.1007/JHEP04(2017)083} {\bibfield  {journal} {\bibinfo
  {journal} {JHEP}\ }\textbf {\bibinfo {volume} {04}}\bibfield  {number}
  {\bibinfo  {number} { (83)},\ \bibinfo {pages} {083}},\ }\Eprint
  {https://arxiv.org/abs/1701.07356} {arXiv:1701.07356 [hep-ph]} \BibitemShut
  {NoStop}%
\bibitem [{\citenamefont {Crisanti}\ \emph
  {et~al.}(2026{\natexlab{a}})\citenamefont {Crisanti}, \citenamefont
  {Lippstreu}, \citenamefont {McLeod},\ and\ \citenamefont
  {Polackova}}]{Crisanti:2026cuf}%
  \BibitemOpen
  \bibfield  {author} {\bibinfo {author} {\bibfnamefont {G.}~\bibnamefont
  {Crisanti}}, \bibinfo {author} {\bibfnamefont {L.}~\bibnamefont {Lippstreu}},
  \bibinfo {author} {\bibfnamefont {A.~J.}\ \bibnamefont {McLeod}},\ and\
  \bibinfo {author} {\bibfnamefont {M.}~\bibnamefont {Polackova}},\ }\href@noop
  {} {\bibinfo {title} {{DiscKosky: Efficient Genealogical Constraints from
  Sector-By-Sector Critical Point Counting}}} (\bibinfo {year}
  {2026}{\natexlab{a}}),\ \Eprint {https://arxiv.org/abs/2610.40772}
  {arXiv:2610.40772 [hep-th]} \BibitemShut {NoStop}%
\bibitem [{\citenamefont {Crisanti}\ \emph
  {et~al.}(2026{\natexlab{b}})\citenamefont {Crisanti}, \citenamefont
  {Frellesvig}, \citenamefont {Pokraka},\ and\ \citenamefont
  {Smith}}]{Crisanti:2026rbc}%
  \BibitemOpen
  \bibfield  {author} {\bibinfo {author} {\bibfnamefont {G.}~\bibnamefont
  {Crisanti}}, \bibinfo {author} {\bibfnamefont {H.}~\bibnamefont
  {Frellesvig}}, \bibinfo {author} {\bibfnamefont {A.}~\bibnamefont
  {Pokraka}},\ and\ \bibinfo {author} {\bibfnamefont {S.}~\bibnamefont
  {Smith}},\ }\bibfield  {title} {\bibinfo {title} {{Magic Relations and
  Critical Varieties of Feynman Integrals}},\ }\href@noop {} {\bibfield
  {journal} {\bibinfo  {journal} {arXiv preprint}\ } (\bibinfo {year}
  {2026}{\natexlab{b}})},\ \Eprint {https://arxiv.org/abs/2605.29789}
  {arXiv:2605.29789 [hep-th]} \BibitemShut {NoStop}%
\bibitem [{\citenamefont {Boels}\ \emph {et~al.}(2016)\citenamefont {Boels},
  \citenamefont {Kniehl},\ and\ \citenamefont {Yang}}]{Boels:2015yna}%
  \BibitemOpen
  \bibfield  {author} {\bibinfo {author} {\bibfnamefont {R.}~\bibnamefont
  {Boels}}, \bibinfo {author} {\bibfnamefont {B.~A.}\ \bibnamefont {Kniehl}},\
  and\ \bibinfo {author} {\bibfnamefont {G.}~\bibnamefont {Yang}},\ }\bibfield
  {title} {\bibinfo {title} {{Master integrals for the four-loop Sudakov form
  factor}},\ }\href {https://doi.org/10.1016/j.nuclphysb.2015.11.016}
  {\bibfield  {journal} {\bibinfo  {journal} {Nucl. Phys. B}\ }\textbf
  {\bibinfo {volume} {902}},\ \bibinfo {pages} {387} (\bibinfo {year}
  {2016})},\ \Eprint {https://arxiv.org/abs/1508.03717} {arXiv:1508.03717
  [hep-th]} \BibitemShut {NoStop}%
\bibitem [{\citenamefont {Tkachov}(1981)}]{Tkachov:1981wb}%
  \BibitemOpen
  \bibfield  {author} {\bibinfo {author} {\bibfnamefont {F.~V.}\ \bibnamefont
  {Tkachov}},\ }\bibfield  {title} {\bibinfo {title} {{A Theorem on Analytical
  Calculability of Four Loop Renormalization Group Functions}},\ }\href
  {https://doi.org/10.1016/0370-2693(81)90288-4} {\bibfield  {journal}
  {\bibinfo  {journal} {Phys. Lett.}\ }\textbf {\bibinfo {volume} {100B}},\
  \bibinfo {pages} {65} (\bibinfo {year} {1981})}\BibitemShut {NoStop}%
\bibitem [{\citenamefont {Chetyrkin}\ and\ \citenamefont
  {Tkachov}(1981)}]{Chetyrkin:1981qh}%
  \BibitemOpen
  \bibfield  {author} {\bibinfo {author} {\bibfnamefont {K.~G.}\ \bibnamefont
  {Chetyrkin}}\ and\ \bibinfo {author} {\bibfnamefont {F.~V.}\ \bibnamefont
  {Tkachov}},\ }\bibfield  {title} {\bibinfo {title} {{Integration by Parts:
  The Algorithm to Calculate beta Functions in 4 Loops}},\ }\href
  {https://doi.org/10.1016/0550-3213(81)90199-1} {\bibfield  {journal}
  {\bibinfo  {journal} {Nucl. Phys. B}\ }\textbf {\bibinfo {volume} {192}},\
  \bibinfo {pages} {159} (\bibinfo {year} {1981})}\BibitemShut {NoStop}%
\bibitem [{\citenamefont {Laporta}(2000)}]{Laporta:2001dd}%
  \BibitemOpen
  \bibfield  {author} {\bibinfo {author} {\bibfnamefont {S.}~\bibnamefont
  {Laporta}},\ }\bibfield  {title} {\bibinfo {title} {{High precision
  calculation of multiloop Feynman integrals by difference equations}},\ }\href
  {https://doi.org/10.1016/S0217-751X(00)00215-7, 10.1142/S0217751X00002157}
  {\bibfield  {journal} {\bibinfo  {journal} {Int. J. Mod. Phys.}\ }\textbf
  {\bibinfo {volume} {A15}},\ \bibinfo {pages} {5087} (\bibinfo {year}
  {2000})},\ \Eprint {https://arxiv.org/abs/hep-ph/0102033}
  {arXiv:hep-ph/0102033 [hep-ph]} \BibitemShut {NoStop}%
\bibitem [{\citenamefont {Henn}\ \emph {et~al.}(2024)\citenamefont {Henn},
  \citenamefont {Matija{\v{s}}i{\'c}}, \citenamefont {Miczajka}, \citenamefont
  {Peraro}, \citenamefont {Xu},\ and\ \citenamefont {Zhang}}]{Henn:2024ngj}%
  \BibitemOpen
  \bibfield  {author} {\bibinfo {author} {\bibfnamefont {J.~M.}\ \bibnamefont
  {Henn}}, \bibinfo {author} {\bibfnamefont {A.}~\bibnamefont
  {Matija{\v{s}}i{\'c}}}, \bibinfo {author} {\bibfnamefont {J.}~\bibnamefont
  {Miczajka}}, \bibinfo {author} {\bibfnamefont {T.}~\bibnamefont {Peraro}},
  \bibinfo {author} {\bibfnamefont {Y.}~\bibnamefont {Xu}},\ and\ \bibinfo
  {author} {\bibfnamefont {Y.}~\bibnamefont {Zhang}},\ }\bibfield  {title}
  {\bibinfo {title} {{A computation of two-loop six-point Feynman integrals in
  dimensional regularization}},\ }\href
  {https://doi.org/10.1007/JHEP08(2024)027} {\bibfield  {journal} {\bibinfo
  {journal} {JHEP}\ }\textbf {\bibinfo {volume} {08}}\bibfield  {number}
  {\bibinfo  {number} { (8)},\ \bibinfo {pages} {027}},\ }\Eprint
  {https://arxiv.org/abs/2403.19742} {arXiv:2403.19742 [hep-ph]} \BibitemShut
  {NoStop}%
\end{thebibliography}%
\end{document}